\documentclass[aps,prd,showpacs,onecolumn,preprintnumbers,notitlepage,
nofootinbib,amsfont,amssymb,10pt,singlespacing] {revtex4-1}
\usepackage{amsmath,bm}
\usepackage{times}
\usepackage{braket}
\usepackage{color,graphicx}
\usepackage{slashed}
\usepackage{hyperref}

\newcommand{\beq}{\begin{eqnarray}}
\newcommand{\eeq}{\end{eqnarray}}
\newcommand{\non}{\nonumber\\ }
\newcommand{\ov}{\overline}

\newcommand{\acp}{{\cal A}_{CP}}

\newcommand{\psl}{ P \hspace{-2.8truemm}/ }

\newcommand{\epsl}{\epsilon \hspace{-1.8truemm}/\,  }

\def\lsim{ {\ \lower-1.2pt\vbox{\hbox{\rlap{$<$}\lower6pt\vbox{\hbox{$\sim$}
}}}\ } }
\def\gsim{ {\ \lower-1.2pt\vbox{\hbox{\rlap{$>$}\lower6pt\vbox{\hbox{$\sim$}
}}}\ } }

\def \jhep{ J. High Energy Phys.  }

\definecolor{Red}{rgb}{1.,0.,0.}

\definecolor{Blue}{rgb}{0.,0.,1.}

\newcommand{\orcid}[1]{\thanks{\href{http://orcid.org/#1}{ORCID: #1}}}

\definecolor{nicered}{rgb}{0.7,0.1,0.2}
\definecolor{nicegreen}{rgb}{0.1,0.4,0.2}

\hypersetup{colorlinks,citecolor=nicegreen,linkcolor=nicered}

\begin{document}
\title{\boldmath $B_{d,s}^0 \to f_1 f_1$ decays with $f_1(1285)-f_1(1420)$ mixing
in the perturbative QCD approach}
\author{Zewen~Jiang$^{1}$}
\author{De-Hua~Yao$^{1}$}
\author{Zhi-Tian~Zou$^{2}$}
\author{Xin~Liu$^{1}$}
\email
[Electronic address: ]
{liuxin@jsnu.edu.cn}
\orcid{0000-0001-9419-7462}
\author{Ying~Li$^{2}$}
\author{Zhen-Jun~Xiao$^{3}$}
\affiliation{
$^1$ Department of Physics,
Jiangsu Normal University, Xuzhou 221116, China
\\
$^2$ Department of Physics, Yantai University, Yantai 264005, China
\\
$^3$ Department of Physics and Institute of Theoretical
Physics, Nanjing Normal University, Nanjing 210023, China
}


\date{\today{}}

\begin{abstract}

We investigate the $B_{d,s}^0 \to f_1 f_1$ decays
in the framework of perturbative QCD(PQCD) approach
with a referenced value $\phi_{f_1} \sim 24^\circ$.
Here, $f_1$ denotes the axial-vector mesons $f_1(1285)$
and $f_1(1420)$ with mixing angle $\phi_{f_1}$ in the
quark-flavor basis. The observables such as branching
ratios, direct {\it CP} violations, and polarization fractions
of the $B_s^0 \to f_1 f_1$ decays are predicted for the first time.
We find that:
(i) the almost pure penguin modes $B_s^0 \to f_1 f_1$
have large branching ratios in the order of $10^{-6}
\sim 10^{-5}$ due to the Cabibbo-Kobayashi-Maskawa
enhancement and generally constructive
interferences between the amplitudes of $B_s^0
\to f_n f_s$ and $B_s^0 \to f_s f_s$ with $f_n$
and $f_s$ being the quark-flavor states of $f_1$ mesons.
(ii) The observables receive important contributions
from the weak annihilation diagrams
in the PQCD approach. In particular, without
the annihilation contributions, the $B_s^0 \to
f_1(1420) f_1(1420)$ branching ratio
will decrease about 81\% and its longitudinal
polarization fraction will reduce around 43\%. And
(iii) the dependence of the $B_{d,s}^0 \to f_1 f_1$
decay rates on $\phi_{f_1}$ exhibits some
interesting line shapes, whose confirmations
would be helpful to constrain the determination
of $\phi_{f_1}$ inversely. All the PQCD predictions
await for the (near) future examinations at
Large Hadron Collider beauty and/or Belle-II
experiments to further understand the
properties of the axial-vector mesons
and the perturbative dynamics released
from the considered decay modes.

\end{abstract}


\preprint{\footnotesize  JSNU-PHY-HEP-02/20}
\maketitle


%
%

\section{Introduction}

As listed in the Particle Data Group(PDG)~\cite{Zyla:2020}, the
$f_1(1285)$ and its partner, namely, the $f_1(1420)$,~\footnote{It
is noted that the $f_1(1420)$ is generally considered as the partner
of the $f_1(1285)$~\cite{Zyla:2020}, although the authors stated that
both the $f_1(1420)$ and the $f_1(1510)$ are partners of the $f_1(1285)$~\cite{Chen:2015iqa}. In this work, we will take the $f_1(1420)$
as the partner of the $f_1(1285)$. For more information about these two
axial-vector states, please refer to the mini review ``63. pseudoscalar and
pseudovector mesons in the 1400 MeV region"~\cite{Zyla:2020} in the PDG2020
for detail, and references therein.} are categorized into the light axial-vector
meson family with a spin-parity quantum number $J^P=1^+$. In the naive quark model, according to the spectroscopic notation $n^{2{S}+1}\!{L}_{J}$ with
radial excitation $n$, spin multiplicity $2{S}+1$, relative angular momentum ${L}$,
and total spin ${J}$~\cite{Amsler:2004ps}, they are one type
of the {\it p}-wave mesons, namely, $1^3\!P_1$. Analogous to $\eta-\eta^{\prime}$
mixing in the pseudoscalar sector~\cite{Zyla:2020}, due to ${\rm SU}(3)$
flavor symmetry breaking effects, these two $f_1$ mesons [for the sake of
simplicity, hereafter, we will use $f_1$ to denote both $f_1(1285)$ and
$f_1(1420)$ unless otherwise stated] also demand the admixtures of the
flavor states $f_n \equiv \frac{u\bar u + d\bar d}{\sqrt{2}}$ and $f_s
\equiv s\bar s$ in the quark-flavor basis and could be described as a
$2\times 2$ rotation matrix~\cite{Aaij:2013rja}:
 \beq
\left(
\begin{array}{c} f_1(1285)\\ f_1(1420) \\ \end{array} \right ) &=&
  \left( \begin{array}{cc}
 \cos{\phi_{f_1}} & -\sin{\phi_{f_1}} \\
 \sin{\phi_{f_1}} & \; \;\ \cos{\phi_{f_1}} \end{array} \right )
 \left( \begin{array}{c}  f_{n}\\ f_{s} \\ \end{array} \right )\;,
 \label{eq:mix-fn-fs}
 \eeq
with a mixing angle $\phi_{f_1}$, which is correlated with the
angle $\theta_{f_1}$ in the singlet-octet basis via the following
relation,
\beq
 \phi_{f_1} &=& \theta_i - \theta_{f_1} \;.
 \label{eq:phi-theta}
\eeq
Here, $\theta_i$ is the ``ideal" mixing angle with the value
$\theta_i = 35.3^\circ$. It is therefore clear to see that $\phi_{f_1}$
could be as a probe to examine the deviation from ideal mixing. On one hand,
the definite understanding of this $\phi_{f_1}$(or $\theta_{f_1}$) could
shed light on the structure of these two $f_1$ mesons; On the other hand,
it is of great interest to note that, as one of the three important
mixing angles in the sector of axial-vector mesons, $\phi_{f_1}$(or
$\theta_{f_1}$) has the potential to help constrain the distinct mixing
between $K_{1A}$ and $K_{1B}$ states with angle $\theta_{K_1}$~\cite{Cheng:2011pb,Zyla:2020}, where the former is a
$^3\!P_1$ state while the latter is a $^1\!P_1$ one. It means that
the good constraints on $\phi_{f_1}$(or $\theta_{f_1}$) could
indirectly pin down the $\theta_{K_1}$ to better investigate the
structure of $K_{1}(1270)$ and $K_1(1400)$
mesons~\cite{Cheng:2007mx,Feldmann:2014iha,Liu:2014dxa,Liu:2014doa,Liu:2014jsa}.

Up to now, there are several explorations on the
$\phi_{f_1}$(or $\theta_{f_1}$) at both theoretical and experimental aspects~\cite{Gidal:1987bn,Close:1997nm,Li:2000dy,Li:2005eq,
Carvalho:2002fh,Yang:2007zt,Cheng:2007mx,Yang:2008xw,Cheng:2008gxa,
Yang:2010ah,Cheng:2011pb,Dudek:2011tt,Stone:2013eaa,Dudek:2013yja,Cheng:2013cwa,
Aaij:2013rja,Liu:2014doa,Close:2015rza}.
One cannot yet determine definitely its value due to limited understanding
on the nature of these two $f_1$ states, although,
about seven years ago, the Large Hadron Collider beauty(LHCb) collaboration
extracted experimentally $\phi_{f_1}=(24.0^{+3.1+0.6}_{-2.6-0.8})^\circ$
with a twofold ambiguity from the $B_{d,s}^0 \to J/\psi f_1(1285)$ decays
for the first time~\cite{Aaij:2013rja}.
Because there are no interferences between the flavor $f_n$ and $f_s$
states in this type of decay modes, this ambiguity is expected to be
settled in the decay modes with significantly constructive or destructive
interferences between those two flavor states,
for example, in the $B_{(s)} \to f_1 P$ decays~\cite{Liu:2014jsa},
the $B_{(s)} \to f_1 V$~\cite{Liu:2016rqu} modes, and other
$B_{(s)}/D_{(s)} \to f_1 M$($M$ stands for the possible mesons)
channels. However, it is worth pointing out
that the $D_{(s)} \to f_1 M$ decays cannot yet be perturbatively
calculated based on the QCD theory. Hence,
those relevant $D_{(s)}$ meson decays have to be left for future studies
elsewhere. In this work, we will study the $B_{d,s}^0 \to f_1 f_1$ decays
in the perturbative QCD(PQCD) approach~\cite{Keum:2000ph} based on the
$k_T$ factorization theorem at leading order~\footnote{To our knowledge,
the ``$\pi\pi, K\pi$" puzzle, e.g.,~\cite{Kpi_puzzle,Liu:2015sra},
in the heavy $B$ meson decays stimulated the development of the
factorization approaches to higher order, representatively, the
next-to-next-to-leading order calculations~\cite{NNLO-QCDF}
in the QCD factorization(QCDF) approach~\cite{Beneke:1999br}.
The PQCD approach has also started its next-to-leading order trip gradually~\cite{Li:2010nn,Liu:2015sra}.
But, in fact, it is well known that, according to the perturbation theory,
the contributions at leading order are usually predominant. }.
The significant interferences among the $B_{d,s}^0 \to f_n f_n$,
$f_n f_s$, and $f_s f_s$ decay amplitudes could be observed
in the considered modes, just like those in the pseudoscalar
$B_{d,s}^0 \to \eta^{(\prime)} \eta^{(\prime)}$ cases~\cite{Xiao:2006mg,Ali:2007ff}.
As discussed in Ref.~\cite{Liu:2014jsa}, due to the consistency between
the latest calculations from Lattice QCD~\cite{Dudek:2013yja}
and the current measurement from LHCb~\cite{Aaij:2013rja}, we will
adopt  $\phi_{f_1} = 24^\circ$
as a referenced value to make quantitative
evaluations and phenomenological discussions.

In the literature, the $B_{d}^0 \to f_1 f_1$ decays have been
investigated in the QCDF approach,
and the decay rates and the longitudinal polarization
fractions have been collected in the Table~X of Ref.~\cite{Cheng:2008gxa}.
However, the predicted branching ratios are too small to be measured
in the near future at LHCb and/or Belle-II experiments.
Compared to these Cabibbo-Kobayashi-Maskawa(CKM) suppressed
$B_d^0 \to f_1 f_1$ modes, the CKM favored $B_s^0 \to f_1 f_1$
ones are expected to be measurable with possibly large decay
rates due to the naive enhancement of $|\frac{V_{ts}}{V_{td}}|^2 \sim 20$
for both penguin-dominated channels or of $|\frac{V_{ts}V_{tb}}{V_{ub}V_{ud}}|^2
\sim 100$ for the penguin-dominated $B_s^0$ while the tree-dominated $B_d^0$
decays, apart from the possibly constructive interferences in the
$B_{s}^0 \to f_1 f_1$ decays. To our best knowledge, the $B_s^0 \to
f_1 f_1$ decays presented in this work are studied theoretically
for the first time in the literature. Moreover, as discussed in
Ref.~\cite{Cheng:2008gxa}, power corrections in QCDF always involve
troublesome end-point divergences. Therefore, more parameters
are introduced to parametrize the contributions arising from
the non-factorizable emission and the annihilation diagrams~\cite{Beneke:2001ev},
which results in large theoretical uncertainties. Objectively speaking,
the QCDF approach is a powerful tool for analyzing the $B$ meson decays
by global fitting to the data. But, the data-fitting and/or model-dependent
parametrization always make it lose the predictive power more or less.

The PQCD approach we adopted in this work is one of the important and popular
factorization methods based on QCD dynamics. It is known that the PQCD approach,
based on the $k_T$ factorization theorem, is free of end-point divergences by keeping
quarks' transverse momentum and the Sudakov formalism makes it more self-consistent.
Thus, the PQCD approach doesn't need to introduce any other parameters, except for
the essential non-perturbative inputs, namely, wave functions or distribution amplitudes
for the initial and final mesons. Note that, these inputs
are universal and are usually computed in the non-perturbative techniques
such as QCD sum rules and Lattice QCD, or extracted from the available
experimental data. A distinct advantage of the PQCD approach
is that one can really do the quantitative calculations of form factor,
non-factorizable emission and annihilation type diagrams,
apart from the factorizable emission ones.
It is worth addressing that one has realized the importance of
annihilation contributions in the heavy flavor $B$ and $D$ meson
decays, for example, the predictions of {\it CP}-violating asymmetries
of $B_d^0 \to \pi^\pm \pi^\mp$, $K^\pm \pi^\mp$ decays~\cite{Keum:2000ph,Hong:2005wj},
the explanations to polarization problem of
$B \to \phi K^*$ modes~\cite{Li:2004mp,Li:2004ti,Gritsan:2007hs}, and the
explorations of phenomenologies of $D^0 \to \pi^\pm \pi^\mp$, $K^\pm K^\mp$
channels~\cite{Li:2012cfa}, and so forth. And what is more, the confirmation
from LHCb experiment on the pure annihilation $B_d^0 \to K^+ K^-$
and $B_s^0 \to \pi^+ \pi^-$ decay rates predicted in the PQCD approach
are very exciting~\cite{Xiao:2011tx,Aaltonen:2011jv}.
Actually, the PQCD predictions for the $B \to PP$, $PV$, and $VV$ decays
have shown good consistency globally with the existing data within errors.
It means that the PQCD approach has the unique advantage and general
reliability at the aspects of calculating the hadronic matrix elements
in the heavy $B$ meson decays. The interested readers could refer
to the review article~\cite{Keum:2000ph} for more details about
this PQCD approach.


\section{ Formalism and perturbative calculations}\label{sec:form}

The decay amplitude for $B_{d,s}^0 \to f_1 f_1$ decays
in the PQCD approach can be conceptually written as follows:
\beq
A(B_{d,s}^0 \to f_1 f_1) &\sim &\int\!\! d x_1 d
x_2 d x_3 b_1 d b_1 b_2 d b_2 b_3 d b_3
\non && \cdot {\mathrm{Tr}}
\left [ C(t) \Phi_{B_{d,s}^0}(x_1, b_1) \Phi_{f_1}(x_2, b_2)
\Phi_{f_1}(x_3, b_3) H(x_i, b_i, t) S_t(x_i)\, e^{-S(t)} \right ]\;,
\label{eq:amp-f1f1}
\eeq
in which, $x_i(i=1,2,3)$ is the momentum fraction of the valence
quark in the initial and final state mesons; $b_i$ is the conjugate
space coordinate of the transverse momentum $k_{iT}$; Tr denotes
the trace over Dirac and SU(3) color indices;
$C(t)$ stands for the Wilson coefficients including
the large logarithms $\ln (m_W/t)$~\cite{Keum:2000ph};
$t$ is the largest running energy scale in hard kernel $H(x_i,b_i,t)$;
and $\Phi$ is the wave function describing the hadronization
of quark and anti-quark to a meson (the explicit form of the involved
wave functions associated with the distribution amplitudes
can be found later in the Appendix~\ref{sec:app1} ). The jet
function $S_t(x_i)$ comes from threshold resummation, which
exhibits a strong suppression effect in the small $x$
region~\cite{Li:2001ay,Li:2002mi}, while
the Sudakov factor $e^{-S(t)}$ arises from $k_T$ resummation,
which provides a strong suppression in the small $k_T$ (or large $b$) region~\cite{Botts:1989kf,Li:1992nu}.
These resummation effects therefore guarantee the removal of the end-point
singularities. The detailed expressions for $S_t(x_i)$ and $e^{-S(t)}$ can
be easily found in Refs.~\cite{Li:2001ay,Li:2002mi,Botts:1989kf,Li:1992nu}.
Note that, to keep the consistency, we will use the leading order
Wilson coefficients in the following calculations.
For the renormalization group evolution of the Wilson coefficients
from higher scale to lower scale, we will adopt the formulas
in Ref.~\cite{Keum:2000ph} directly.


For the $B_{d,s}^0 \to f_1 f_1$ decays, the related weak effective
Hamiltonian $H_{{\rm eff}}$ can be read as~\cite{Buchalla:1995vs}
\beq
H_{\rm eff}\, &=&\, {G_F\over\sqrt{2}}
\biggl\{ V^*_{ub}V_{uq} \biggl[ C_1(\mu)O_1^{u}(\mu)
+C_2(\mu)O_2^{u}(\mu) \biggr]
 - V^*_{tb}V_{tq} \biggl[ \sum_{i=3}^{10}C_i(\mu)O_i(\mu) \biggr] \biggr\}+ {\rm H.c.}\;,
\label{eq:heff}
\eeq
in which, $q=d\ {\rm or}\ s $, $G_F=1.16639\times 10^{-5}{\rm
GeV}^{-2}$ is the  Fermi constant, $V$ denotes the CKM matrix elements,
and $C_i(\mu)$ stands for Wilson coefficients at the renormalization scale
$\mu$. The local four-quark operators $O_i(i=1,\cdots,10)$ are written as
\begin{enumerate}
\item[]{(1) Tree
operators}
\begin{eqnarray}
{\renewcommand\arraystretch{1.5}
\begin{array}{ll}
\displaystyle
O_1^{u}\, =\,
(\bar{q}_\alpha u_\beta)_{V-A}(\bar{u}_\beta b_\alpha)_{V-A}\;,
& \displaystyle
O_2^{u}\, =\, (\bar{q}_\alpha u_\alpha)_{V-A}(\bar{u}_\beta b_\beta)_{V-A}\;;
\end{array}}
\label{eq:operators-1}
\end{eqnarray}

\item[]{(2) QCD penguin operators}
\begin{eqnarray}
{\renewcommand\arraystretch{1.5}
\begin{array}{ll}
\displaystyle
O_3\, =\, (\bar{q}_\alpha b_\alpha)_{V-A}\sum_{q'}(\bar{q}'_\beta q'_\beta)_{V-A}\;,
& \displaystyle
O_4\, =\, (\bar{q}_\alpha b_\beta)_{V-A}\sum_{q'}(\bar{q}'_\beta q'_\alpha)_{V-A}\;,
\\
\displaystyle
O_5\, =\, (\bar{q}_\alpha b_\alpha)_{V-A}\sum_{q'}(\bar{q}'_\beta q'_\beta)_{V+A}\;,
& \displaystyle
O_6\, =\, (\bar{q}_\alpha b_\beta)_{V-A}\sum_{q'}(\bar{q}'_\beta q'_\alpha)_{V+A}\;;
\end{array}}
\label{eq:operators-2}
\end{eqnarray}

\item[]{(3) Electroweak penguin operators}
\begin{eqnarray}
{\renewcommand\arraystretch{1.5}
\begin{array}{ll}
\displaystyle
O_7\, =\,
\frac{3}{2}(\bar{q}_\alpha b_\alpha)_{V-A}\sum_{q'}e_{q'}(\bar{q}'_\beta q'_\beta)_{V+A}\;,
& \displaystyle
O_8\, =\,
\frac{3}{2}(\bar{q}_\alpha b_\beta)_{V-A}\sum_{q'}e_{q'}(\bar{q}'_\beta q'_\alpha)_{V+A}\;,
\\
\displaystyle
O_9\, =\,
\frac{3}{2}(\bar{q}_\alpha b_\alpha)_{V-A}\sum_{q'}e_{q'}(\bar{q}'_\beta q'_\beta)_{V-A}\;,
& \displaystyle
O_{10}\, =\,
\frac{3}{2}(\bar{q}_\alpha b_\beta)_{V-A}\sum_{q'}e_{q'}(\bar{q}'_\beta q'_\alpha)_{V-A}\;,
\end{array}}
\label{eq:operators-3}
\end{eqnarray}
\end{enumerate}
with the color indices $\alpha, \ \beta$ and the notations
$(\bar{q}'q')_{V\pm A} = \bar q' \gamma_\mu (1\pm \gamma_5)q'$.
The index $q'$ in the summation of the above operators runs
through $u,\;d,\;s$, $c$, and $b$.

\begin{figure}[!!htb]
\centering
\begin{tabular}{l}
\includegraphics[width=0.8\textwidth]{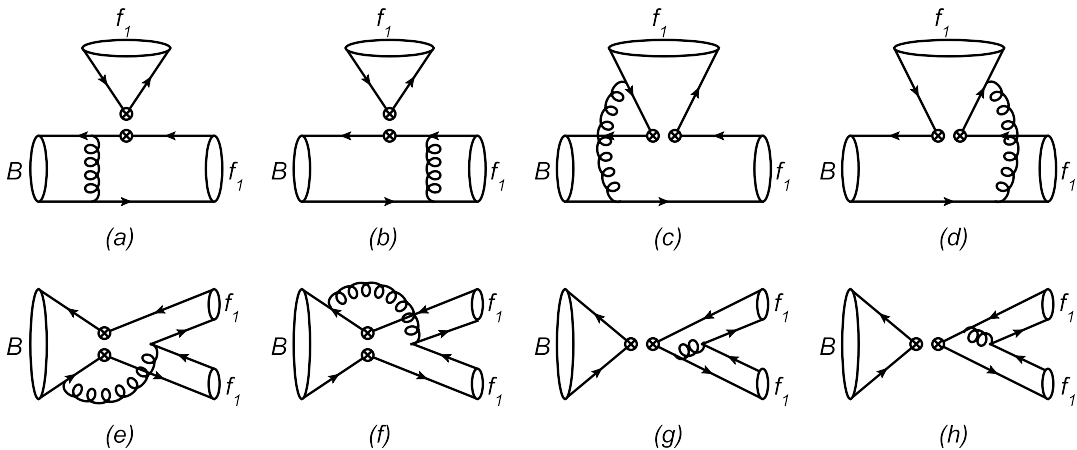}
\end{tabular}
\caption{Typical Feynman diagrams contributing to the
$B_{d,s}^0 \to f_1 f_1$ decays in the PQCD approach
at leading order. Here, $B$ and $f_1$ stand for the initial
$B_{d}^0$ and $B_s^0$ and the final $f_1(1285)$ and
$f_1(1420)$ mesons, respectively. }
  \label{fig:fig1}
\end{figure}

As illustrated in Fig.~\ref{fig:fig1}, it is easy to find that
the considered $B_{d,s}^0 \to f_1 f_1$ decays contain two kinds
of topologies of the diagrams, namely, the emission one and
the annihilation one, which include eight types of diagrams
in the PQCD approach at leading order:
(i) factorizable [non-factorizable] emission diagrams Figs.~\ref{fig:fig1}(a)
and~\ref{fig:fig1}(b) [Figs.~\ref{fig:fig1}(c) and~\ref{fig:fig1}(d)]
in the first row; and (ii) non-factorizable [factorizable] annihilation diagrams
Figs.~\ref{fig:fig1}(e) and~\ref{fig:fig1}(f) [Figs.~\ref{fig:fig1}(g)
and~\ref{fig:fig1}(h)] in the second row, respectively.
With the effective Hamiltonian and various operators as shown in
Eqs.~(\ref{eq:heff})-(\ref{eq:operators-3}),
we can straightforwardly calculate the contributions in the PQCD approach.
Hereafter, for the sake of simplicity, we will adopt
$F$ and $F^{P_1}$ ($M$ and $M^{P_1}$) to denote the factorizable
(non-factorizable) Feynman amplitudes induced by the $(V-A)(V-A)$
and $(V-A)(V+A)$ operators, and $F^{P_2}$ ($M^{P_2}$) to denote the
factorizable (non-factorizable) Feynman amplitudes from the $(S-P)(S+P)$
operators, which are resulted from a Fierz transformation of the $(V-A)(V+A)$ ones.

A remark is in order for the Feynman amplitudes: the Feynman amplitudes
for the $B$ meson decaying into two axial-vector mesons have been
collected in~\cite{Liu:2012jb}. In this work,
it is not necessary for us to list the same calculations
existed in the literature. The interested readers could
refer to Eqs.~(25)-(60)~\cite{Liu:2012jb} for detail.
In Ref.~\cite{Zou:2015iwa}, the authors studied the $B_{(s)} \to
VV$ decays by keeping the higher power terms proportional
to $r^2_V=m_V^2/m_B^2$ in the denominator of propagators for
virtual quarks and gluons, which resulted in the predictions
for most branching ratios and polarization fractions in the PQCD
approach being in good agreement with the existing measurements.
In light of this success, we would like to retain the terms proportional to
$r_{f_1}^2=m_{f_1}^2/m_{B_{d,s}^0}^2$ in the $B_{d,s}^0 \to f_1 f_1$ decays too.
In fact, we have also taken this strategy into account in the
studies of $B \to f_1 V$ decays~\cite{Liu:2016rqu}.

Together with various contributions from different diagrams as
presented in Eqs.~(25)-(60)~\cite{Liu:2012jb} and the quark-flavor
mixing scheme as shown in Eq.~(\ref{eq:mix-fn-fs}), the decay amplitudes
of six $B_{d,s}^0 \to f_1 f_1$ channels can thus be written
in terms of the combinations of $B_{d,s}^0 \to f_n f_n, f_n f_s,$ and
$f_s f_s$ with different coefficients as follows [the superscript $h$ in
the following formulas stands for the helicity amplitudes with longitudinal($L$),
normal($N$), and transverse($T$) polarizations, respectively]:
\begin{itemize}

\item[1.]{For $B_d^0 \to f_1 f_1$ decays,}

The decay amplitudes for the $B_d^0$ meson decaying into the flavor states
$f_n f_n$, $f_n f_s$, and $f_s f_s$
can be easily written as follows:
\beq
2 A_h(B_d^0 \to f_n f_n) &=&
V_{ub}^* V_{ud} \biggl\{ a_2 (f_{f_n} F^h_{fe}+ f_{B_d^0} F^h_{fa} )
 + C_2 (M^h_{nfe} + M^h_{nfa} ) \biggr\}
 \non &&
- V_{tb}^* V_{td} \biggl\{\biggl[2 a_3 +a_4- 2a_5 -\frac{1}{2}(a_7- a_{9}+ a_{10})\biggr]
f_{f_n} F^h_{fe}
\non &&
+ \biggl[2 a_3+a_4+ 2a_5 +\frac{1}{2}(a_7
+ a_{9}- a_{10})\biggr] f_{B_d^0} F^h_{fa}
+ (a_6 - \frac{1}{2} a_8)f_{B_d^0} F_{fa}^{h,P_2}
\non &&
+\biggl[C_3 + 2 C_4 - \frac{1}{2} (C_9
-C_{10})\biggr] (M^h_{nfe} + M^h_{nfa}
 )\non &&
+(C_5 -\frac{1}{2} C_7) (M_{nfe}^{h,P_1} + M_{nfa}^{h,P_1}
)+(2 C_6 + \frac{1}{2} C_8) (M_{nfe}^{h,P_2}
+ M_{nfa}^{h,P_2} )\biggr\}\;,
\label{eq:fnfn-d}
\eeq
\beq
\sqrt{2}A_h(B_d^0 \to f_n f_s)&=&
-V_{tb}^* V_{td} \biggl\{ \biggl[a_3 -a_5 + \frac{1}{2}
(a_7 - a_9)\biggr] f_{f_{s}} F^h_{fe}
+ (C_4 -\frac{1}{2} C_{10})
{M}^h_{nfe} +(C_6 - \frac{1}{2} C_8)
{M}_{nfe}^{h,P_2} \biggr\}\;,
\label{eq:fnfs-d}
\eeq
\beq
A_h(B_d^0 \to f_s f_s)&=&
-V_{tb}^* V_{td} \biggl\{ \biggl[a_3+a_5- \frac{1}{2} (a_7 + a_9)\biggr]
f_{B_d^0} F^h_{fa}
+(C_4 -\frac{1}{2} C_{10}) M^h_{nfa}
+(C_6 - \frac{1}{2} C_8)
M_{nfa}^{h,P_2} \biggr\}\;.
\label{eq:fsfs-d}
\eeq
In the above formulas, i.e., Eqs.~(\ref{eq:fnfn-d})-(\ref{eq:fsfs-d}),
the subscripts ``$(n)fe$" and ``$(n)fa$" are the abbreviations of
(non-)factorizable emission and (non-)factorizable
annihilation, and $a_i$ is the standard combination
of the Wilson coefficients $C_i$ defined as follows:
\beq
a_1 &=& C_2 + \frac{C_1}{3}\;,\qquad
a_2 = C_1 + \frac{C_2}{3}\;;
\label{eq:Wilson-a-t}
\eeq
\beq
a_i&=& \left\{ \begin{array}{ll}
C_i + C_{i+1}/3 \;\;\;\;\; (i=3,5,7,9),&  \\
C_i + C_{i-1}/3 \;\;\;\;\; (i=4,6,8,10).&   \\ \end{array} \right.
\label{eq:Wilson-a-p}
\eeq
where $C_2 \sim 1$ is
the largest one among all the Wilson coefficients.

The decay amplitudes for the physical states are then
\beq
A_h(B_d^0 \to f_1(1285) f_1(1420))
&=&\sin(2\phi_{f_1})\biggl[A_h(B_d^0 \to f_n f_n)-A_h(B_d^0 \to f_s f_s)\biggr]+\cos(2\phi_{f_1})A_h(B_d^0 \to f_n f_s)\;,
 \label{eq:amp-24-d}
\eeq
\beq
\sqrt{2}A_h(B_d^0 \to f_1(1285) f_1(1285))
&=& 2\cos^2\phi_{f_1} A_h(B_d^0 \to f_n f_n)
+2\sin^2\phi_{f_1} A_h(B_d^0 \to f_s f_s)
\non &&
-\sin(2\phi_{f_1})A_h(B_d^0 \to f_n f_s)\;,
 \label{eq:amp-22-d}
\eeq
\beq
\sqrt{2}A_h(B_d^0 \to f_1(1420) f_1(1420))
&=& 2\sin^2\phi_{f_1} A_h(B_d^0 \to f_n f_n)
+2\cos^2\phi_{f_1} A_h(B_d^0 \to f_s f_s)
\non &&
 +\sin(2\phi_{f_1})A_h(B_d^0 \to f_n f_s)\;.
 \label{eq:amp-44-d}
\eeq

\item[2.]{For $B_s^0 \to f_1 f_1$ decays}

Analogously, the decay amplitudes of $B_s^0 \to f_n f_n$, $f_n f_s$,
and $f_s f_s$ can be written as,
\beq
2A_h(B_s^0 \to f_n f_n)&=&
V_{ub}^*V_{us}\biggl\{  a_2 f_{B_s^0} F^h_{fa}
 + C_2 M^h_{nfa} \biggr\}\non &&
- V_{tb}^*V_{ts} \biggl\{
 \biggl[2(a_3+a_5)+\frac{1}{2}(a_7+a_9)\biggr] f_{B_s^0} F^h_{fa}
\non && +(2C_4 + \frac{1}{2} C_{10})
M^h_{nfa}
+ (2C_6 + \frac{1}{2} C_{8}) M_{nfa}^{h,P_2}
  \biggr\}\;,
  \label{eq:fnfn-s}
\eeq
\beq
\sqrt{2}A_h(B_s^0 \to f_n f_s)&=&V_{ub}^*V_{us}
\biggl\{ a_2 f_{f_n} F^h_{fe}+C_2 M^h_{nfe}
\biggr\}
\non &&
- V_{tb}^*V_{ts} \biggl\{ \biggl[ 2(a_3 -a_5) - \frac{1}{2}(a_7 - a_9)\biggr]
f_{f_n} F^h_{fe}
\non &&
+(2C_4 + \frac{1}{2} C_{10}) M^h_{nfe}
+ (2C_6 + \frac{1}{2} C_8) M_{nfe}^{h,P_2}\biggr\} \;,
\label{eq:fnfs-s}
\eeq
\beq
A_h(B_s^0 \to f_s f_s)&=& -V_{tb}^*V_{ts}\biggl\{
\biggl[ a_3+a_4-a_5+\frac{1}{2}
(a_7-a_9-a_{10})\biggr]
f_{f_s} F^h_{fe}
\non &&
+\biggl[a_3+a_4+a_5-\frac{1}{2}(a_7+a_9+a_{10}) \biggr]
f_{B_s^0} F^h_{fa}
\non &&
+(a_6-\frac{1}{2}a_8) f_{B_s^0} F_{fa}^{h,P_2} +\biggl[C_3 +C_4-\frac{1}{2}(C_9+C_{10})\biggr]
\non && \times
( M^h_{nfe} +M^h_{nfa} )
+(C_5 - \frac{1}{2} C_7) (M_{nfe}^{h,P_1} +M_{nfa}^{h,P_1} )
\non &&
+(C_6 - \frac{1}{2} C_8) (M_{nfe}^{h,P_2} +M_{nfa}^{h,P_2} ) \biggr\}\;.
\label{eq:fsfs-s}
\eeq
Then, we could give the decay amplitudes for the physical states similarly,
\beq
A_h(B_s^0 \to f_1(1285)f_1(1420))
&=& \sin(2\phi_{f_1})\biggl[A_h(B_s^0 \to f_n f_n) - A_h(B_s^0 \to f_s f_s)\biggr]+
\cos(2\phi_{f_1})A_h(B_s^0 \to f_n f_s)\;,
\label{eq:amp-24-s}
\eeq
\beq
\sqrt{2}A_h(B_s^0 \to f_1(1285) f_1(1285))
&=& 2\cos^2\phi_{f_1} A_h(B_s^0 \to f_n f_n) -\sin(2\phi_{f_1})A_h(B_s^0 \to f_n f_s)
\non &&
+2\sin^2\phi_{f_1} A_h(B_s^0 \to f_s f_s)\;,
\label{eq:amp-22-s}
\eeq
\beq
\sqrt{2}A_h(B_s^0 \to f_1(1420) f_1(1420))
&=& 2\sin^2\phi_{f_1} A_h(B_s^0 \to f_n f_n) +\sin(2\phi_{f_1})A_h(B_s^0 \to f_n f_s)
\non &&
+2 \cos^2\phi_{f_1} A_h(B_s^0 \to f_s f_s) \;.
\label{eq:amp-44-s}
\eeq
\end{itemize}

\section{Numerical Results and Discussions} \label{sec:randd}

Now, we will perform the numerical calculations in the PQCD approach
on the experimental observables such as the {\it CP}-averaged branching
ratios(${\cal B}$), the direct {\it CP}-violating asymmetries($\acp^{\rm dir}$),
and the {\it CP}-averaged
polarization fractions, etc. for the considered $B_{d,s}^0 \to f_1 f_1$ decays.
Some essential comments on the input parameters are in order:
\begin{itemize}
\item[]{(a) Distribution amplitudes for the flavor states $f_n$ and $f_s$}

As discussed in~\cite{Yang:2007zt}, the $^3\!P_1$-axial-vector meson has the
similar behavior to the vector one. Meanwhile, it is noted that,
for the distribution amplitudes, the flavor $\eta_n$
and $\eta_s$ states of $\eta^{(\prime)}$ usually took the same form as pion but with
different decay constants $f_{\eta_n}$ and $f_{\eta_s}$ in the pseudoscalar sector.
Therefore, for the flavor states $f_n$ and $f_s$ in this work, we shall adopt the
the same distribution amplitudes as those of the $a_1(1260)$ meson. The decay
constants $f_{f_n}$ and $f_{f_s}$, and the relevant Gegenbauer moments
can be easily found in Refs.~\cite{Yang:2007zt,Li:2009tx,Verma:2011yw,Liu:2014doa,Liu:2016rqu}.

\item[]{(b) Wolfenstein parametrization of CKM matrix and four parameters}

For the CKM matrix elements, we adopt the Wolfenstein
parametrization at leading order~\cite{Wolfenstein:1983yz}
and the updated parameters released in PDG2018:
$A=0.836$, $\lambda=0.22453$, $\bar{\rho}=0.122^{+0.018}_{-0.017}$,
and $\bar{\eta}=0.355^{+0.012}_{-0.011}$~\cite{Tanabashi:2018oca}.

\item[]{(c) QCD scale, masses, and $B_{d,s}^0$ meson lifetimes}

The relevant QCD scale~({\rm GeV}), masses~({\rm GeV}),
and $B$ meson lifetime({\rm ps})
are the following
~\cite{Keum:2000ph,Yang:2007zt,Verma:2011yw,Aaij:2013rja,Tanabashi:2018oca}
\beq
 \Lambda_{\overline{\rm{MS}}}^{(f=4)} &=& 0.250\; , \quad m_W = 80.41\;,
 \quad  m_{B_d^0}= 5.28\;, \quad  m_{B_s^0}= 5.37\;, \quad  m_b = 4.8 \;; \non
  f_{f_{n}}&=& 0.193^{+0.043}_{-0.038}\;,
  \quad f_{f_{s}} = 0.230 \pm 0.009\;,
\quad m_{f_{n}}= 1.28\;,
\quad m_{f_{s}}= 1.42\;;\\
  \tau_{B_d^0}&=& 1.520\;,
   \quad  \tau_{B_s^0}= 1.509\;, \quad \phi_{f_1} = (24.0^{+3.2}_{-2.7})^\circ\;.
\nonumber
\label{eq:inputs}
\eeq
Of course, in numerical calculations, central values of the above input parameters will be used implicitly unless otherwise stated.
\end{itemize}

\subsection{{\it CP}-averaged branching ratios}
Similar to the $B \to f_1 V$ decays~\cite{Liu:2016rqu},
the $B_{d,s}^0 \to f_1 f_1$ decay rate can also be written as
\beq
\Gamma =\frac{G_{F}^{2}|\bf{P_c}|}{16 \pi m^{2}_{B_{d,s}^0} }
\sum_{h=L,N,T} A_h^{\dagger }  A_h\;
\label{dr1}
\eeq
where $|\bf{P_c}|\equiv |\bf{P_{2z}}|=|\bf{P_{3z}}|$ is the momentum of either the
outgoing axial-vector meson and $A_h$ can be found, for example,
in Eqs.~(\ref{eq:amp-24-s})-(\ref{eq:amp-44-s}). The corresponding branching ratios
${\cal B}$ can thus be easily obtained through the relation
${\cal B} = \tau_{B_{d,s}^0} \Gamma$.

\begin{table}[htb]
\caption{ Theoretical predictions on the quantities of the $B_{d,s}^0 \to f_1(1285) f_1(1420)$ decays obtained in the PQCD approach,
where the errors are sequentially from
the shape parameter $\omega_B$, the decay constants $f_M$, the Gegenbauer moments $a_{f_1}$, the mixing angle $\phi_{f_1}$, the higher order corrections factor $a_t$, and the CKM parameters $V$. }
\label{tab:f12f14-ds}
 \begin{center}\vspace{-0.3cm}{
\begin{tabular}[t]{c|c||c|c}
\hline
   \multicolumn{2}{c||}{Decay Modes}   &  $B_d^0 \to f_1(1285) f_1(1420)$
   &  $B_s^0 \to f_1(1285) f_1(1420)$  \\
\hline \hline
  ${\cal B}$
  & $\Gamma/ \Gamma_{\rm total}$
  &$5.05^{+1.14+4.73+3.00+0.23+1.21+0.11}_{-0.87-2.47-1.92-0.26-0.73-0.09}
 \times 10^{-7} $
  &$ 7.50^{+1.56+1.26+6.25+1.08+1.97+0.01}_{-1.03-1.10-4.16-0.70-1.31-0.01}
 \times 10^{-6} $
 \\
 \hline
 $f_L(\%)$      & $|{\cal A}_L|^2$
 &$18.0^{+0.8+4.8+13.0+0.8+2.2+0.8}_{-0.4-2.5-5.4-0.6-0.4-0.6}
  $
 &$59.0^{+6.4+3.4+2.6+6.0+2.2+0.0}_{-6.6-5.3-4.9-8.4-3.2-0.0}
  $
 \\
 $f_{||}(\%)$   & $|{\cal A}_{||}|^2$
 &$45.3^{+0.3+1.5+3.5+0.3+0.5+0.4}_{-0.4-2.4-7.8-0.2-1.3-0.3}
  $
 &$23.2^{+3.8+3.0+2.8+4.8+1.9+0.1}_{-3.6-1.9-1.5-3.4-1.2-0.0}
  $
  \\
 $f_{\perp}(\%)$& $|{\cal A}_\perp|^2$
 &$ 36.6^{+0.2+1.2+2.0+0.5+0.0+0.4}_{-0.3-2.2-5.0-0.4-0.7-0.4}
$
 &$17.7^{+2.9+2.3+2.2+3.6+1.4+0.0}_{-2.7-1.4-1.2-2.5-0.9-0.0}
  $
 \\
 \hline
 $\phi_{||}$(rad)& $\arg\frac{{\cal A}_{||}}{{\cal A}_L}$
 &$1.89^{+1.40+2.00+2.62+0.27+0.14+2.15}_{-0.00-0.00-0.00-0.00-0.00-0.00}
$
 &$1.99^{+0.16+0.22+0.11+0.22+0.08+0.01}_{-0.17-0.29-0.21-0.24-0.05-0.00}
  $
 \\
 $\phi_{\perp}$(rad)& $\arg\frac{{\cal A}_{\perp}}{{\cal A}_L}$
 &$1.92^{+1.40+2.00+1.54+1.37+0.98+0.07}_{-0.00-0.00-0.10-0.00-0.00-0.07}
  $
 &$2.00^{+0.16+0.25+0.11+0.23+0.08+0.00}_{-0.16-0.29-0.20-0.24-0.05-0.00}
  $
  \\
  \hline
 $\acp^{\rm dir}(\%)$& $\frac{\overline{\Gamma}-\Gamma}{\overline{\Gamma}+\Gamma}$
 &$54.7^{+0.9+1.5+2.8+2.5+0.5+1.4}_{-1.4-3.1-8.6-2.7-2.8-1.3}
  $
 &$2.8^{+0.2+0.0+0.8+0.0+0.0+0.1}_{-0.3-0.3-0.7-0.3-0.0-0.1}
  $
 \\
 $\acp^{\rm dir}(L)(\%)$& $\frac{\bar{f}_L-f_L}{\bar{f}_L+f_L}$
 &$94.8^{+0.0+0.7+2.6+0.0+1.1+1.2}_{-1.4-10.2-23.9-2.0-11.9-1.4}
  $
 &$-5.0^{+0.3+1.1+1.4+0.5+0.0+0.1}_{-0.3-1.5-2.8-0.5-0.2-0.2}
  $
 \\
 $\acp^{\rm dir}(||)(\%)$& $\frac{\bar{f}_{||}-f_{||}}{\bar{f}_{||}+f_{||}}$
 &$46.3^{+0.7+3.4+5.7+3.6+2.4+1.5}_{-1.1-4.2-5.7-3.5-3.5-1.6}
  $
 &$13.6^{+1.9+3.3+0.2+3.1+1.0+0.5}_{-1.9-3.6-1.1-3.7-1.2-0.4}
  $
 \\
 $\acp^{\rm dir}(\perp)(\%)$& $\frac{\bar{f}_\perp-f_\perp}{\bar{f}_\perp+f_\perp}$
 &$45.5^{+0.8+3.0+3.0+3.3+2.2+1.5}_{-1.2-3.7-4.6-3.2-3.4-1.6}
$
 &$14.8^{+1.9+3.6+0.9+3.3+1.2+0.5}_{-2.0-4.0-1.5-4.1-1.4-0.5}
  $
 \\ \hline
\end{tabular}}
\end{center}
\end{table}

The numerical results predicted in the PQCD approach
for the observables, specifically, branching ratios,
direct {\it CP} violations, and polarization fractions
associated with the theoretical errors are collected in
Tables~\ref{tab:f12f14-ds}-\ref{tab:f14f14-ds}.
As for the errors, they are mainly induced by
the uncertainties of the shape parameter
$\omega_B = 0.40 \pm 0.04\ (\omega_B = 0.50 \pm 0.05)$~GeV
in the $B^{0}_d\ (B_s^0)$ meson distribution amplitude,
of the combined
decay constants $f_{M}$ from the $^3\!P_1$-axial-vector state
as $f_{f_{n}}=0.193^{+0.043}_{-0.038}$~GeV and
$f_{f_{s}}=0.230 \pm 0.009$~GeV,
of the combined Gegenbauer moments $a_{f_1}$ from
$a_2^{\parallel}$ and $a_1^{\perp}$
in the $f_{n}$ and $f_{s}$ state distribution amplitudes,
of the mixing angle $\phi_{f_1} = (24.0^{+3.2}_{-2.7})^\circ$ for
the $f_1(1285)-f_1(1420)$ mixing system in the quark-flavor basis,
of the maximal running hard scale $t_{\rm max}=(1.0\pm 0.2)\ t$
\footnote{As mentioned above, parts of the next-to-leading order
corrections to two-body hadronic $B$ meson decays have been
proposed in the PQCD approach~\cite{Li:2010nn,Liu:2015sra}, however,
the higher order QCD contributions
to the decays of $B$ mesons into two vector final states
beyond leading order are not yet available now.
Therefore, the higher order contributions in this work are simply
investigated by exploring
the variation of hard scale $t_{\rm max}$ with 20\%,
i.e., from $0.8t$ to $1.2t$
(not changing $1/b_i, i= 1,2,3$), in the hard kernel, which have
been counted into one of the sources of theoretical uncertainties. As can be
seen in the Tables~\ref{tab:f12f14-ds}-\ref{tab:f14f14-ds}, it looks like that,
relative to the color-suppressed, tree-dominated $B_d^0 \to f_1(1285) f_1(1285)$ mode,
all the other five decays considered in this work are more sensitive to the higher order
corrections potentially.},
and of the combined CKM matrix elements $V$ from the
parameters $\bar \rho$ and $\bar \eta$, respectively.
Note that the errors induced by the hadronic parameters such as the
decay constants and the Gegenbauer moments in
the adopted distribution amplitudes, particularly for the axial-vector states,
are larger than those from other inputs, which can be easily seen from the
Tables~\ref{tab:f12f14-ds}-\ref{tab:f14f14-ds}. Frankly speaking, due to the
lack of the essential constraints from experiments, we have to choose
the available parameters calculated in the QCD sum rules with large uncertainties.
Therefore, it is expected that the experimental examinations on
the numerical results and the theoretical predictions presented in this work
could provide effective constraints on these hadronic parameters in the (near) future.
Meanwhile, the calculations of the above-mentioned inputs
arising from lattice QCD could also help better understand the relevant hadron
dynamics and give more precise predictions theoretically.

Based on the effective Hamiltonian as shown in Eq.~(\ref{eq:heff}),
it is clear to see that,  at the quark level, the $B_{d}^0 \to f_1 f_1$
decays are the $\Delta S = 0$ (here, the capital $S$ describes the strange
flavor number)
type modes with the $\bar b \to \bar d$ transition, while the
$B_s^0 \to f_1 f_1$ ones are the  $\Delta S = 1$ type channels with the
$\bar b \to \bar s$ transition, where the former is CKM suppressed,
and the latter is, however,
CKM favored. Then, as generally expected, the $B_s^0 \to f_1 f_1$ decay rates
are much larger than the $B_d^0 \to f_1 f_1$ ones with different extents
due to the CKM enhancement
and the constructive/destructive interferences among the flavor
$f_n f_n$, $f_n f_s$, and $f_s f_s$ final states.
The ${\cal B}(B_{d,s}^0 \to f_1 f_1)$ predicted in the PQCD approach
can confirm this expectation numerically. One
can see the predictions as presented explicitly in the Tables~\ref{tab:f12f14-ds}-\ref{tab:f14f14-ds}.
Within a bit large theoretical uncertainties, the branching ratios of $B_{d,s}^0 \to f_1 f_1$ decays
in the PQCD approach can be read as follows,
\beq
{\cal B}(B_d^0 \to f_1(1285) f_1(1285)) &=& 6.64^{+8.70}_{-4.25} \times 10^{-7}\;, \qquad
{\cal B}(B_s^0 \to f_1(1285) f_1(1285)) = 3.70^{+3.49}_{-2.39} \times 10^{-6}\;,
\label{eq:BR-22-ds}\\
{\cal B}(B_d^0 \to f_1(1285) f_1(1420)) &=& 5.05^{+5.85}_{-3.34} \times 10^{-7}\;, \qquad
{\cal B}(B_s^0 \to f_1(1285) f_1(1420)) = 7.50^{+6.94}_{-4.67} \times 10^{-6}\;,
\label{eq:BR-24-ds}\\
{\cal B}(B_d^0 \to f_1(1420) f_1(1420)) &=& 1.00^{+1.05}_{-0.62} \times 10^{-7}\;, \qquad
{\cal B}(B_s^0 \to f_1(1420) f_1(1420)) = 3.37^{+3.27}_{-2.18} \times 10^{-5}\;.
\label{eq:BR-44-ds}
\eeq
where all the errors arising from the input parameters have been added in quadrature.

\begin{table}[htb]
\caption{ Same as Table~\ref{tab:f12f14-ds} but for $B_{d,s}^0 \to f_1(1285) f_1(1285)$ decays. }
\label{tab:f12f12-ds}
 \begin{center}\vspace{-0.3cm}{
\begin{tabular}[t]{c|c||c|c}
\hline
   \multicolumn{2}{c||}{Decay Modes}   &  $B_d^0 \to f_1(1285) f_1(1285)$
   &  $B_s^0 \to f_1(1285) f_1(1285)$  \\
\hline \hline
  ${\cal B}$
  & $\Gamma/ \Gamma_{\rm total}$
  &$6.64^{+1.21+8.12+2.68+0.51+0.80+0.35}_{-0.96-3.82-1.40-0.60-0.38-0.31}
 \times 10^{-7} $
  &$3.70^{+1.30+2.15+2.11+0.70+0.95+0.02}_{-0.90-1.34-1.55-0.54-0.66-0.01}
 \times 10^{-6} $
 \\
 \hline
 $f_L(\%)$      & $|{\cal A}_L|^2$
 &$27.7^{+3.4+3.7+18.2+2.3+5.0+0.4}_{-3.4-2.5-9.9-1.7-3.4-0.4}
  $
 &$
78.9^{+1.7+1.5+4.6+1.1+0.0+0.1}_{-1.8-1.7-7.0-1.2-0.1-0.2}
  $
 \\
 $f_{||}(\%)$   & $|{\cal A}_{||}|^2$
 &$38.2^{+1.9+1.8+6.0+0.9+2.2+0.2}_{-1.9-2.0-10.3-1.3-2.9-0.2}
  $
 &$11.7^{+1.0+0.9+3.8+0.6+0.1+0.1}_{-1.0-0.8-2.6-0.6-0.0-0.1}
  $
  \\
 $f_{\perp}(\%)$& $|{\cal A}_\perp|^2$
 &$34.1^{+1.5+1.1+4.0+0.7+1.2+0.2}_{-1.5-1.7-8.0-1.1-2.1-0.2}
$
 &$9.4^{+0.8+0.8+3.3+0.5+0.1+0.1}_{-0.7-0.7-2.1-0.5-0.0-0.0}
  $
 \\
 \hline
 $\phi_{||}$(rad)& $\arg\frac{{\cal A}_{||}}{{\cal A}_L}$
 &$4.10^{+0.02+0.04+0.11+0.03+0.17+0.03}_{-0.03-0.06-1.38-0.04-0.09-0.02}
$
 &$4.00^{+0.15+0.25+0.16+0.15+0.05+0.00}_{-0.13-0.16-0.13-0.11-0.02-0.00}
  $
 \\
 $\phi_{\perp}$(rad)& $\arg\frac{{\cal A}_{\perp}}{{\cal A}_L}$
 &$4.11^{+0.02+0.05+0.11+0.04+0.18+0.03}_{-0.02-0.07-1.41-0.04-0.09-0.02}
  $
 &$4.02^{+0.15+0.24+0.16+0.15+0.04+0.00}_{-0.13-0.16-0.13-0.11-0.03-0.00}
  $
  \\
  \hline
 $\acp^{\rm dir}(\%)$& $\frac{\overline{\Gamma}-\Gamma}{\overline{\Gamma}+\Gamma}$
 &$26.5^{+0.7+5.1+10.8+3.5+1.7+0.9}_{-0.8-7.3-17.7-4.5-2.5-0.9}
  $
 &$-0.3^{+0.5+0.7+0.8+0.5+0.2+0.0}_{-0.4-0.5-1.1-0.3-0.1-0.0}
  $
 \\
 $\acp^{\rm dir}(L)(\%)$& $\frac{\bar{f}_L-f_L}{\bar{f}_L+f_L}$
 &$-72.0^{+7.8+5.2+28.1+3.3+17.1+2.4}_{-8.3-4.0-19.5-2.8-17.6-2.2}
  $
 &$-2.8^{+0.2+0.3+0.9+0.2+0.4+0.1}_{-0.0-0.2-0.9-0.1-0.3-0.0}
  $
 \\
 $\acp^{\rm dir}(||)(\%)$& $\frac{\bar{f}_{||}-f_{||}}{\bar{f}_{||}+f_{||}}$
 &$66.3^{+4.3+0.4+0.0+0.4+1.5+2.3}_{-4.7-1.4-9.9-0.8-1.7-2.5}
  $
 &$9.1^{+3.1+5.3+2.6+3.4+0.5+0.3}_{-2.6-4.0-3.7-2.7-0.3-0.3}
  $
 \\
 $\acp^{\rm dir}(\perp)(\%)$& $\frac{\bar{f}_\perp-f_\perp}{\bar{f}_\perp+f_\perp}$
 &$62.0^{+4.1+0.6+1.1+0.4+2.0+2.3}_{-4.3-1.3-10.3-0.7-2.0-2.1}
$
 &$8.6^{+3.1+5.7+2.6+3.6+0.7+0.3}_{-2.6-4.2-3.6-2.8-0.4-0.3}
  $
 \\ \hline
\end{tabular}}
\end{center}
\end{table}

As aforementioned, the $B_d^0 \to f_1 f_1$ decays have been investigated
in the QCDF approach~\cite{Cheng:2008gxa}. The numerical results with
large errors presented in~\cite{Cheng:2008gxa} can be read as
follows~\footnote{As discussed in~\cite{Liu:2014jsa}, the predictions in
the QCDF approach for the $B \to f_1 M$ decays~\cite{Cheng:2007mx,Cheng:2008gxa}
with $M$ being the pseudoscalar, vector, and axial-vector mesons provided in
the second entry could be quoted to make effective comparisons to those given
in the quark-flavor basis, i.e., Eq.~(\ref{eq:mix-fn-fs}) with a positive angle
in the PQCD approach.}:
 \beq
{\cal B}(B_d^0 \to f_1(1285) f_1(1285)) &=& 0.2^{+0.2+2.5}_{-0.1-0.0}
\times 10^{-6}\;,
\label{eq:BR-22-d-QCDF}\\
{\cal B}(B_d^0 \to f_1(1285) f_1(1420)) &=& 0.05^{+0.05+0.63}_{-0.00-0.00}
\times 10^{-6}\;,
\label{eq:BR-24-d-QCDF}\\
{\cal B}(B_d^0 \to f_1(1420) f_1(1420)) &=& 0.01^{+0.01+0.06}_{-0.00-0.00}
\times 10^{-6}\;.
\label{eq:BR-44-d-QCDF}
\eeq
The largest errors are from the parametrized hard spectator
scattering and annihilation diagrams, as mentioned in~\cite{Cheng:2008gxa}.
Note that the parametrization of these contributions are inferred from those
in the $B \to VV$ decays in the QCDF approach
due to the similar behavior between the vector meson and the
$^3\!P_1$ axial-vector one. One can see that the $B_d^0 \to f_1 f_1$
decay rates predicted in the PQCD and QCDF approaches are roughly
consistent with each other
within large uncertainties, although, in terms of the central values, the
branching ratios of the latter two modes in the QCDF approach are smaller than
those in the PQCD approach with one order.
It is worth pointing out that the dramatically different central values of these
$B_d^0$ decays, especially the latter two modes, by an order of magnitude in the QCDF and
PQCD approaches maybe mainly resulted from the different hard scales, that is, the largest
running scale $\mu=t_{\rm max}$ in the PQCD approach, while the fixed hard scale $\mu=m_b$ in
the QCDF approach,
and from
the different treatments on the hard spectator interactions and the annihilation diagrams,
that is, those contributions are quantitatively calculated in the PQCD approach, while they are roughly parametrized in the QCDF approach due to endpoint singularity.

\begin{table}[htb]
\caption{ Same as Table~\ref{tab:f12f14-ds} but for $B_{d,s}^0 \to f_1(1420) f_1(1420)$ decays. }
\label{tab:f14f14-ds}
 \begin{center}\vspace{-0.3cm}{
\begin{tabular}[t]{c|c||c|c}
\hline
   \multicolumn{2}{c||}{Decay Modes}   &  $B_d^0 \to f_1(1420) f_1(1420)$
   &  $B_s^0 \to f_1(1420) f_1(1420)$  \\
\hline \hline
  ${\cal B}$
  & $\Gamma/ \Gamma_{\rm total}$
  &$1.00^{+0.20+0.75+0.59+0.33+0.23+0.00}_{-0.14-0.42-0.33-0.23-0.15-0.00}
 \times 10^{-7} $
  &$3.37^{+0.29+0.58+3.04+0.12+1.01+0.01}_{-0.24-0.50-2.00-0.17-0.65-0.00}
 \times 10^{-5} $
 \\
 \hline
 $f_L(\%)$      & $|{\cal A}_L|^2$
 &$12.4^{+5.1+5.4+19.5+2.7+3.4+0.2}_{-3.9-1.9-6.0-1.5-2.8-0.2}
  $
 &$10.6^{+1.6+1.5+9.8+0.9+1.9+0.0}_{-1.8-1.4-3.3-1.0-1.8-0.0}
  $
 \\
 $f_{||}(\%)$   & $|{\cal A}_{||}|^2$
 &$49.7^{+2.3+0.9+3.9+0.8+1.8+0.2}_{-3.0-3.1-11.7-1.5-2.1-0.2}
  $
 &$51.1^{+1.0+0.7+2.2+0.5+1.0+0.0}_{-0.9-0.9-6.4-0.6-1.1-0.0}
  $
  \\
 $f_{\perp}(\%)$& $|{\cal A}_\perp|^2$
 &$37.9^{+1.7+0.9+2.2+0.7+1.1+0.1}_{-2.1-2.4-7.7-1.2-1.3-0.0}
$
 &$38.3^{+0.9+0.7+1.1+0.5+0.7+0.0}_{-0.7-0.6-3.4-0.3-0.7-0.0}
  $
 \\
 \hline
 $\phi_{||}$(rad)& $\arg\frac{{\cal A}_{||}}{{\cal A}_L}$
 &$4.22^{+0.20+0.10+0.34+0.08+0.22+0.00}_{-0.14-0.18-1.25-0.11-0.16-0.01}
$
 &$3.37^{+0.00+0.09+0.05+0.06+0.13+0.00}_{-0.01-0.09-0.06-0.06-0.10-0.00}
  $
 \\
 $\phi_{\perp}$(rad)& $\arg\frac{{\cal A}_{\perp}}{{\cal A}_L}$
 &$4.25^{+0.20+0.10+0.34+0.08+0.22+0.00}_{-0.11-0.18-1.25-0.11-0.16-0.01}
  $
 &$3.40^{+0.00+0.09+0.05+0.06+0.13+0.00}_{-0.01-0.09-0.06-0.06-0.10-0.00}
  $
  \\
  \hline
 $\acp^{\rm dir}(\%)$& $\frac{\overline{\Gamma}-\Gamma}{\overline{\Gamma}+\Gamma}$
 &$25.4^{+1.8+6.9+5.0+5.0+3.2+0.8}_{-2.2-8.3-11.9-4.9-3.0-0.7}
  $
 &$-1.9^{+0.2+0.3+0.4+0.2+0.2+0.1}_{-0.2-0.3-0.6-0.2-0.2-0.1}
  $
 \\
 $\acp^{\rm dir}(L)(\%)$& $\frac{\bar{f}_L-f_L}{\bar{f}_L+f_L}$
 &$9.4^{+19.1+25.9+39.7+18.3+11.9+0.4}_{-9.4-20.6-67.8-14.0-17.7-0.4}
  $
 &$2.3^{+0.6+0.1+0.8+0.1+0.7+0.0}_{-0.5-0.4-0.6-0.2-0.7-0.1}
  $
 \\
 $\acp^{\rm dir}(||)(\%)$& $\frac{\bar{f}_{||}-f_{||}}{\bar{f}_{||}+f_{||}}$
 &$28.0^{+0.4+4.4+5.1+3.0+2.3+0.9}_{-0.6-4.5-4.0-2.7-3.1-0.8}
  $
 &$-2.3^{+0.2+0.3+0.5+0.2+0.2+0.0}_{-0.3-0.4-1.2-0.3-0.2-0.1}
  $
 \\
 $\acp^{\rm dir}(\perp)(\%)$& $\frac{\bar{f}_\perp-f_\perp}{\bar{f}_\perp+f_\perp}$
 &$27.3^{+0.4+4.3+3.0+3.0+2.3+0.9}_{-0.6-4.6-3.3-2.7-3.2-0.8}
$
 &$-2.5^{+0.2+0.4+0.6+0.2+0.3+0.1}_{-0.2-0.3-1.1-0.2-0.2-0.0}
  $
 \\ \hline
\end{tabular}}
\end{center}
\end{table}

\begin{table}[hbt]
\caption{Decay amplitudes(in units of $10^{-3}\; \rm{GeV}^3$) of the $B_d^0 \to f_1 f_1$ modes with three polarizations in the PQCD approach,
where only the central values are quoted
for clarification. Note that the numerical results in the parentheses are the corresponding amplitudes without annihilation contributions. }
\label{tab:DecAmp-d-f1f1}
 \begin{center}\vspace{-0.3cm}{\footnotesize
\begin{tabular}[t]{c||c|c||c|c||c|c}
\hline  \hline
   Decay Modes   &  \multicolumn{2}{c||}{$B_d^0 \to f_1(1285) f_1(1285)$} &  \multicolumn{2}{c||}{$B_d^0 \to f_1(1285) f_1(1420)$} &  \multicolumn{2}{c}{$B_d^0 \to f_1(1420) f_1(1420)$} \\
   \hline
   Contributions & Tree diagrams
   & Penguin diagrams
 & Tree diagrams & Penguin diagrams  & Tree diagrams & Penguin diagrams  \\
 \hline \hline
 $A_L$
 &$\begin{array}{c}0.463 - i\; 0.005 \;
 \\ (0.299 - i\; 0.133)\end{array}$
 &$\begin{array}{c}0.253 - i\; 0.306 \;
 \\ (0.219 - i\; 0.056)\end{array}$
 &$\begin{array}{c}0.292 - i\; 0.003 \;
 \\ (0.189 - i\; 0.084)\end{array}$
 &$\begin{array}{c}-0.285 + i\; 0.102 \;
 \\ (-0.198 + i\; 0.124)\end{array}$
 &$\begin{array}{c}0.092 - i\; 0.001 \;
 \\ (0.059 - i\; 0.026)\end{array}$
 &$\begin{array}{c} -0.024 - i\; 0.132 \;
 \\ (-0.168 + i\; 0.089)\end{array}$
 \\
 \hline
$A_N$
 &$\begin{array}{c}-0.109 - i\; 0.441 \;
 \\ (-0.116 - i\; 0.447)\end{array}$
 &$\begin{array}{c}-0.047 + i\; 0.203 \;
 \\ (-0.024 + i\; 0.050)\end{array}$
 &$\begin{array}{c}-0.069 - i\; 0.278 \;
 \\ (-0.073 - i\; 0.281)\end{array}$
 &$\begin{array}{c}-0.278 + i\; 0.282 \;
 \\ (-0.262 + i\; 0.184)\end{array}$
 &$\begin{array}{c}-0.022 - i\; 0.087 \;
 \\ (-0.023 - i\; 0.089)\end{array}$
 &$\begin{array}{c} -0.162 + i\; 0.134 \;
 \\ (-0.160 + i\; 0.106)\end{array}$
  \\
 \hline
$A_T$
 &$\begin{array}{c}-0.236 - i\; 0.964 \;
 \\ (-0.228 - i\; 0.964)\end{array}$
 &$\begin{array}{c}-0.114 + i\; 0.413 \;
 \\ (-0.065 + i\; 0.102)\end{array}$
 &$\begin{array}{c}-0.149 - i\; 0.607 \;
 \\ (-0.143 - i\; 0.607)\end{array}$
 &$\begin{array}{c}-0.581 + i\; 0.565 \;
 \\ (-0.549 + i\; 0.366)\end{array}$
 &$\begin{array}{c}-0.047 - i\; 0.191 \;
 \\ (-0.045 - i\; 0.191)\end{array}$
 &$\begin{array}{c} -0.341 + i\; 0.267 \;
 \\ (-0.333 + i\; 0.210)\end{array}$
 \\
 \hline \hline
\end{tabular}}
\end{center}
\end{table}

According to the mixing scheme in Eq.~(\ref{eq:mix-fn-fs}) with referenced value
$\phi_{f_1} =24^\circ$, one can easily find that the $f_1(1285)$
is predominated by the $f_n$ component, while the $f_1(1420)$ is
governed by the $f_s$ component. Hence, for the $B_d^0 \to f_1 f_1$ decays,
it could be naively anticipated that the $B_d^0 \to f_1(1285) f_1(1285)\
[B_d^0 \to f_1(1420) f_1(1420)]$ mode is tree- (penguin-) diagram dominant,
but with only a few percent
of penguin (tree) contaminations. For the $B_d^0 \to f_1(1285) f_1(1420)$
channel, both of the tree diagrams and the penguin ones contribute evidently to
the decay rate simultaneously.
The decay amplitudes induced by the tree diagrams and the penguin diagrams
for the $B_d^0 \to f_1 f_1$ decays have been collected and can be seen clearly
in the Table~\ref{tab:DecAmp-d-f1f1}. To clarify the above expectations,
we present the {\it CP}-averaged branching ratios in the PQCD approach
without considering the tree contributions for the considered
$B_d^0 \to f_1 f_1$ decays, namely, ${\cal B}(B_d^0 \to f_1(1285) f_1(1285)) = 1.63 \times 10^{-7}$, ${\cal B}(B_d^0 \to f_1(1285) f_1(1420)) = 3.29 \times 10^{-7}$,
and ${\cal B}(B_d^0 \to f_1(1420) f_1(1420)) = 0.86 \times 10^{-7}$ with about
75\%, 35\%, and 15\% reduction, respectively. Here, only the central values
are quoted for clarifications.

In principle, the $B_d^0 \to f_1 f_1$ decay rates
could be accessible at the LHCb and/or Belle-II experiments with
accumulating a large number of $B_d^0\bar B_d^0$ events in the near future,
after all, the $B_d^0 \to K^+ K^-$ with decay rate $1.3 \pm 0.5 \times 10^{-7}$
and $B_s^0 \to \pi^+ \pi^-$ with branching ratio
$7.6 \pm 1.9 \times 10^{-7}$~\cite{Aaltonen:2011jv,Amhis:2019ckw,Zyla:2020}
have been measured at LHCb.
But, by considering the secondary
decay process of $f_1(1285)$, namely,
${\cal B}(f_1(1285) \to \eta \pi^+ \pi^-) \sim 35\%$~\cite{Zyla:2020} or
${\cal B}(f_1(1285) \to 2\pi^0 \pi^+ \pi^-) \sim 22.3\%$~\cite{Zyla:2020}, then
the $B_d^0 \to f_1(1285) f_1(1285)$ channel have to be detected
through the processes $B_d^0 \to f_1(1285)(\to \eta \pi^+ \pi^-) f_1(1285)
(\to \eta \pi^+ \pi^-)$ or $B_d^0 \to f_1(1285)(\to 2\pi^0 \pi^+\pi^-) f_1(1285)
(\to 2\pi^0 \pi^+ \pi^-)$ with the branching ratios under the narrow width approximation,
\beq
{\cal B}(B_d^0 \to (\eta \pi^+ \pi^-)_{f_1(1285)} (\eta \pi^+ \pi^-)_{f_1(1285)})&\equiv& {\cal B}(B_d^0 \to f_1(1285) f_1(1285))
\cdot {\cal B}^2(f_1(1285) \to \eta \pi^+ \pi^+) \non
&\approx& 0.81^{+1.07}_{-0.52} \times 10^{-7}\;,
\eeq
\beq
{\cal B}(B_d^0 \to (2 \pi^0 \pi^+\pi^-)_{f_1(1285)} (2 \pi^0 \pi^+ \pi^-)_{f_1(1285)})
&\equiv& {\cal B}(B_d^0 \to f_1(1285) f_1(1285))
\cdot {\cal B}^2(f_1(1285) \to 2\pi^0  \pi^+ \pi^-) \non
&\approx& 0.33^{+0.43}_{-0.21} \times 10^{-7}\;.
\eeq
It seems that the above two results are too small to be
measured experimentally in the near future.

However, the $B_s^0 \to f_1 f_1$ decays have large branching ratios in the
order of $10^{-6} \sim 10^{-5}$, which
are expected to be measured in the near future
at LHCb and Belle-II experiments.
Unlike the $B_d^0 \to f_1 f_1$ decays, the $B_s^0 \to f_1 f_1$ ones
are almost dominated by the pure penguin contributions just with the tiny while
negligible tree pollution, which can be clearly seen from the decay amplitudes
presented in the Table~\ref{tab:DecAmp-s-f1f1}. Furthermore,
when the contributions from tree diagrams are turned off for the $B_s^0 \to
f_1 f_1$ decays, the {\it CP}-averaged branching ratios, in terms of the central values, will change slightly as follows,
\beq
{\cal B}(B_s^0 \to f_1(1285) f_1(1285)) &=& 3.78 \times 10^{-6}\;,
\label{eq:BR-22-nt-s}\\
{\cal B}(B_s^0 \to f_1(1285) f_1(1420)) &=& 7.43 \times 10^{-6}\;,
\label{eq:BR-24-nt-s}\\
{\cal B}(B_s^0 \to f_1(1420) f_1(1420)) &=& 3.40 \times 10^{-5}\;.
\label{eq:BR-44-nt-s}
\eeq
Relative to the dominant $f_1(1285) \to \eta \pi\pi$ decay,
the decay rate for the dominant $f_1(1420) \to K \bar K \pi$ process is
not yet available~\footnote{Due to the currently unknown nature~\cite{Zyla:2020}
and the different understanding~\cite{Xie:202007} of the $f_1(1420)$,
we just take the absolutely dominant mode, i.e.,
$f_1(1420) \to \bar K K^*$ into account.
Then, by including the secondary decay chain $K^* \to K\pi$ under the narrow
width approximation, the branching ratio of the strong decay
$f_1(1420) \to \bar K K^* \to K_S^0 K^\pm \pi^\mp$ could be naively estimated as
${\cal B}(f_1(1420) \to \bar K K^*) \cdot {\cal B}(K^* \to K^\pm \pi^\mp)$. }.
Therefore, it is not easy to exactly estimate the branching ratios
of $B_s^0 \to f_1(1285) f_1(1420)$ and $f_1(1420) f_1(1420)$ decays via
the resonant channels $B_s^0 \to (\eta\pi^+\pi^-)_{f_1(1285)} (K_S^0 K^+ \pi^-)_{f_1(1420)}$ and $B_s^0 \to (K_S^0 K^+ \pi^- )_{f_1(1420)} (K_S^0
K^+ \pi^{-})_{f_1(1420)}$.
Fortunately, as discussed in Ref.~\cite{Barberis:1997vf},
the only decay modes of the $f_1(1420)$ were assumed as $\bar K K^*$,
$a_0(980) \pi$, and $\phi \gamma$, and the decay rate of $f_1(1420)
\to K^* \bar K$ was given as about 96\%, then the branching ratio
could be naively assumed as ${\cal B}(f_1(1420) \to K_S^0
K^\pm \pi^\mp) \approx 64\%$.
Then, similarly, under the narrow width approximation,
the $B_s^0 \to f_1(1285)(\to \eta \pi^+ \pi^-) f_1(1420)(\to K_S^0 K^+ \pi^-)$
and $B_s^0 \to f_1(1420)(\to K_S^0 K^+ \pi^-) f_1(1420)(\to K_S^0 K^+ \pi^-)$
processes have the branching ratios as follows,
\beq
{\cal B}(B_s^0 \to (\eta\pi^+\pi^-)_{f_1(1285)} (K_S^0 K^+ \pi^-)_{f_1(1420)})&\equiv&
{\cal B}(B_s^0 \to f_1(1420) f_1(1420))\cdot {\cal B}(f_1(1285) \to \eta \pi^+ \pi^-)\non
&&\cdot {\cal B}(f_1(1420) \to K_S^0 K^+ \pi^-)
\approx 1.68^{+1.56}_{-1.04} \times 10^{-6}\;,
\label{eq:24-s-kkpi}
\eeq
\beq
{\cal B}(B_s^0 \to (K_S^0 K^+ \pi^-)_{f_1(1420)} (K_S^0 K^+ \pi^-)_{f_1(1420)})&\equiv&
{\cal B}(B_s^0 \to f_1(1420) f_1(1420))\cdot {\cal B}^2(f_1(1420) \to K_S^0 K^+ \pi^-)\non
&\approx& 1.38^{+1.34}_{-0.89} \times 10^{-5}\;.
\label{eq:44-s-kkpi}
\eeq
Certainly, these two large values as given in Eqs.~(\ref{eq:24-s-kkpi}) and
~(\ref{eq:44-s-kkpi}) are believed to be detectable at LHCb experiments,
as well as at Belle-II ones in the near future.

\begin{table}[hbt]
\caption{Same as Table~\ref{tab:DecAmp-d-f1f1} but for the $B_s^0 \to f_1 f_1$ modes.}
\label{tab:DecAmp-s-f1f1}
 \begin{center}\vspace{-0.3cm}{\footnotesize
\begin{tabular}[t]{c||c|c||c|c||c|c}
\hline  \hline
   Decay Modes   &  \multicolumn{2}{c||}{$B_s^0 \to f_1(1285) f_1(1285)$} &  \multicolumn{2}{c||}{$B_s^0 \to f_1(1285) f_1(1420)$} &  \multicolumn{2}{c}{$B_s^0 \to f_1(1420) f_1(1420)$} \\
   \hline
   Contributions & Tree diagrams
   & Penguin diagrams
 & Tree diagrams & Penguin diagrams  & Tree diagrams & Penguin diagrams  \\
 \hline \hline
 $A_L$
 &$\begin{array}{c}-0.003 + i\; 0.048 \;
 \\ (-0.051 + i\; 0.020)\end{array}$
 &$\begin{array}{c}-2.202 + i\; 0.976 \;
 \\ (-1.530 - i\; 0.060)\end{array}$
 &$\begin{array}{c}0.096 - i\; 0.007 \;
 \\ (0.065 - i\; 0.025)\end{array}$
 &$\begin{array}{c} 2.784 - i\; 1.006 \;
 \\ (2.679 + i\; 0.221)\end{array}$
 &$\begin{array}{c}0.061 - i\; 0.014 \;
 \\ (0.051 - i\; 0.020)\end{array}$
 &$\begin{array}{c} -1.598 + i\; 2.199 \;
 \\ (-0.792 - i\; 0.400)\end{array}$
 \\
 \hline
 $A_N$
 &$\begin{array}{c}0.023 + i\; 0.074 \;
 \\ (0.021 + i\; 0.073)\end{array}$
 &$\begin{array}{c}-0.612 - i\; 0.279 \;
 \\ (-0.891 + i\; 0.117)\end{array}$
 &$\begin{array}{c}-0.026 - i\; 0.091 \;
 \\ (-0.027 - i\; 0.092)\end{array}$
 &$\begin{array}{c} -0.055 + i\; 1.318 \;
 \\ (0.868 + i\; 0.043)\end{array}$
 &$\begin{array}{c}-0.021 - i\; 0.072 \;
 \\ (-0.021 - i\; 0.073)\end{array}$
 &$\begin{array}{c} 3.193 - i\; 2.745 \;
 \\ (1.739 - i\; 0.726)\end{array}$
  \\
 \hline
 $A_T$
 &$\begin{array}{c}0.040 + i\; 0.158 \;
 \\ (0.042 + i\; 0.159)\end{array}$
 &$\begin{array}{c}-1.241 - i\; 0.592 \;
 \\ (-1.827 + i\; 0.209)\end{array}$
 &$\begin{array}{c}-0.055 - i\; 0.202 \;
 \\ (-0.053 - i\; 0.202)\end{array}$
 &$\begin{array}{c} -0.140 + i\; 2.663 \;
 \\ (1.723 + i\; 0.095)\end{array}$
 &$\begin{array}{c}-0.042 - i\; 0.159 \;
 \\ (-0.042 - i\; 0.159)\end{array}$
 &$\begin{array}{c} 6.695 - i\; 5.424 \;
 \\ (3.732 - i\; 1.353)\end{array}$
 \\
 \hline \hline
\end{tabular}}
\end{center}
\end{table}

Different from the ideal mixing between $\omega$ and $\phi$
in the vector sector, both of $f_1(1285)$ and $f_1(1420)$ mesons have
some admixtures of $f_s$ and $f_n$ correspondingly. Therefore,
though the similarity of the distribution amplitudes between the $f_1$ states
and the $\omega(\phi)$ mesons has been observed~\cite{Cheng:2008gxa}, relative to the
$B_{d,s}^0 \to \omega(\phi)\omega(\phi)$ and $B_{d,s}^0 \to f_1 \omega(\phi)$ decays,
the more complicated interferences among the $B_{d,s}^0 \to f_n f_n, f_n f_s,$ and $f_s f_s$ are involved in the $B_{d,s}^0 \to f_1 f_1$ decays, as presented explicitly in the Eqs.~(\ref{eq:amp-24-d})-(\ref{eq:amp-44-d}) and Eqs.~(\ref{eq:amp-24-s})-(\ref{eq:amp-44-s}).
In other words, it is not easy to
naively anticipate the constructive or destructive interferences in the
$B_{d,s}^0 \to f_1 f_1$ decays just like those in the $B_{d,s}^0 \to \omega(\phi)\omega(\phi)$, and $B_{d,s}^0 \to f_1 \omega(\phi)$ ones.
But, as observed in the Table~\ref{tab:DecAmp-fnfs-ds}, the $B_s^0 \to f_n f_n$
channel has a small longitudinal while two tiny and negligible transverse
amplitudes, which would make the interferences in the $B_s^0 \to f_1 f_1$ decays
more easy to be explored.
Therefore, it can still be expected that the nearly pure penguin $B_s^0 \to
f_1(1420) f_1(1420)$ decay with a large branching ratio
could provide useful information to constrain the $B_s^0-\bar B_s^0$ mixing phase,
even to find new physics signal beyond the standard model complementarily.

\begin{table}[hbt]
\caption{Decay amplitudes(in units of $10^{-3}\; \rm{GeV}^3$) for flavor states
$B_{d,s}^0 \to f_n f_n, f_n f_s,$ and $f_s f_s$ of the $B_{d,s}^0 \to f_1 f_1$ modes with three polarizations in the PQCD approach,
where only the central values are quoted
for clarification. Note that the numerical results in the parentheses
are the corresponding amplitudes of
$\bar B_{d,s}^0 \to f_n f_n, f_n f_s,$ and $f_s f_s$.}
\label{tab:DecAmp-fnfs-ds}
 \begin{center}\vspace{-0.3cm}{\footnotesize
\begin{tabular}[t]{c||c|c|c||c|c|c}
\hline  \hline
 Decays &  \multicolumn{3}{c||}{$B_d^0 \to f_1 f_1$} &  \multicolumn{3}{c}{$B_s^0 \to f_1 f_1$} \\
   \hline
   Flavor states & $B_d^0 \to f_n f_n$
   & $B_d^0 \to f_n f_s$
 & $B_d^0 \to f_s f_s$ & $B_s^0 \to f_n f_n$  & $B_s^0 \to f_n f_s$ & $B_s^0 \to f_s f_s$  \\
 \hline \hline
 $A_L$
 &$\begin{array}{c} 0.433 - i\; 0.162 \;
 \\ (-0.176 - i\; 0.315) \end{array}$
 &$\begin{array}{c} -0.336 + i\; 0.159 \;
 \\ (-0.350 - i\; 0.125) \end{array}$
 &$\begin{array}{c} 0.121 - i\; 0.151 \;
 \\ (0.193 - i\; 0.021) \end{array}$
 &$\begin{array}{c} -0.411 + i\; 0.484 \;
 \\ (-0.470 +i\; 0.416) \end{array}$
 &$\begin{array}{c} 2.278 - i\; 0.068 \;
 \\ (2.080 -i\; 0.059) \end{array}$
 &$\begin{array}{c} -2.235 + i\; 1.786 \;
 \\ (-2.235 +i\; 1.786) \end{array}$
 \\
 \hline
 $A_N$
 &$\begin{array}{c} -0.242 - i\; 0.133 \;
 \\ (-0.427 + i\; 0.418) \end{array}$
 &$\begin{array}{c} -0.246 + i\; 0.152 \;
 \\ (-0.282 - i\; 0.067) \end{array}$
 &$\begin{array}{c} 0.002-i\; 0.002 \;
 \\ (0.003 + i\; 0.000) \end{array}$
 &$\begin{array}{c} -0.006 + i\; 0.005 \;
 \\ (-0.008 +i\; 0.002) \end{array}$
 &$\begin{array}{c} 1.923 - i\; 0.553 \;
 \\ (1.912 -i\; 0.280) \end{array}$
 &$\begin{array}{c} 1.833 - i\; 2.142 \;
 \\ (1.833 - i\; 2.142) \end{array}$
  \\
 \hline
 $A_T$
 &$\begin{array}{c} -0.523 - i\; 0.332 \;
 \\ (-0.902 + i\; 0.887) \end{array}$
 &$\begin{array}{c} -0.509 + i\; 0.302 \;
 \\ (-0.573 - i\; 0.147) \end{array}$
 &$\begin{array}{c} 0.001 - i\; 0.004 \;
 \\ (0.004 - i\; 0.002) \end{array}$
 &$\begin{array}{c} -0.003 + i\; 0.006 \;
 \\ (0.000 +i\; 0.007) \end{array}$
 &$\begin{array}{c} 3.997 - i\; 1.059 \;
 \\ (3.959 -i\; 0.468) \end{array}$
 &$\begin{array}{c} 3.858 - i\; 4.259 \;
 \\ (3.858 - i\; 4.259) \end{array}$
 \\
 \hline \hline
\end{tabular}}
\end{center}
\end{table}

As clearly seen in the Tables~\ref{tab:f12f14-ds}-\ref{tab:f14f14-ds},
the numerical
results calculated in the PQCD approach suffer from large
uncertainties induced by the less constrained distribution
amplitudes of the involved hadrons.
At the same time, frankly speaking, it is worthy of stressing that
the large errors presented in the Tables~\ref{tab:f12f14-ds}-\ref{tab:f14f14-ds} for the
considered $B_{d,s}^0 \to f_1 f_1$ decays are mainly induced by the decay constants,
Gegenbauer moments, even the mixing angle of the $f_1$ mesons.
In light of these large uncertainties, we then define some interesting ratios of the branching ratios for the decay modes.
As generally expected, if the modes in a ratio
have similar dependence on a specific input parameter, the error induced by the uncertainty of this input parameter
will be largely canceled in the ratio, even if one cannot make an explicit factorization for this parameter. Furthermore, from the experimental side, we know that the ratios
of the branching ratios generally could be measured
with a better accuracy than that for the individual branching ratios.
The relevant ratios about the decay rates of the considered $B_{d,s}^0 \to
f_1 f_1$ decays can be read as follows:
\beq
R^{sd}_{1}[f_1(1285)f_1(1420)] &\equiv& \frac{{\cal B}(B_s^0 \to f_1(1285) f_1(1420))_{\rm PQCD}}{{\cal B}(B_d^0 \to f_1(1285) f_1(1420))_{\rm PQCD}} \non
&\approx& 14.85^{+0.63}_{-0.21}(\omega_B)^{+9.96}_{-5.89}(f_{M})
^{+2.23}_{-4.18}(a_{f_1})^{+1.40}_{-0.65}(\phi_{f_1})
^{+0.28}_{-0.52}(a_t)^{+0.26}_{-0.30}(V) \;,
\eeq
\beq
R^{sd}_2[f_1(1285)f_1(1285)] &\equiv& \frac{{\cal B}(B_s^0 \to f_1(1285) f_1(1285))_{\rm PQCD}}{{\cal B}(B_d^0 \to f_1(1285) f_1(1285))_{\rm PQCD}} \non
&\approx& 5.57^{+0.80}_{-0.64}(\omega_B)^{+2.80}_{-1.61}(f_{M})
^{+0.66}_{-1.47}(a_{f_1})^{+0.58}_{-0.34}(\phi_{f_1})
^{+0.68}_{-0.71}(a_t)^{+0.26}_{-0.25}(V) \;,
\eeq
\beq
R^{sd}_3[f_1(1420)f_1(1420)] &\equiv& \frac{{\cal B}(B_s^0 \to f_1(1420) f_1(1420))_{\rm PQCD}}{{\cal B}(B_d^0 \to f_1(1420) f_1(1420))_{\rm PQCD}} \non
&\approx& 3.37^{+0.27}_{-0.32}(\omega_B)^{+1.58}_{-1.11}(f_{M})
^{+0.66}_{-1.33}(a_{f_1})^{+0.79}_{-0.75}(\phi_{f_1})
^{+0.19}_{-0.17}(a_t)^{+0.01}_{-0.00}(V)\times 10^{2}\;,
\eeq
\beq
 R^{dd}_4[f_1(1285)/f_1(1420)]  &\equiv& \frac{{\cal B}(B_d^0 \to f_1(1285) f_1(1285))_{\rm PQCD}}{{\cal B}(B_d^0 \to f_1(1285) f_1(1420))_{\rm PQCD}}  \non
 &\approx&  1.31^{+0.05}_{-0.04}(\omega_B)^{+0.20}_{-0.22}(f_{M})
^{+0.36}_{-0.15}(a_{f_1})^{+0.04}_{-0.05}(\phi_{f_1})
^{+0.14}_{-0.12}(a_t)^{+0.04}_{-0.05}(V)\;,
\eeq
\beq
R^{ss}_5[f_1(1420)/f_1(1285)] &\equiv& \frac{{\cal B}(B_s^0 \to f_1(1285) f_1(1420))_{\rm PQCD}}{{\cal B}(B_s^0 \to f_1(1285) f_1(1285))_{\rm PQCD}} \non
&\approx& 2.03^{+0.28}_{-0.22}(\omega_B)^{+0.68}_{-0.53}(f_{M})
^{+0.34}_{-0.48}(a_{f_1})^{+0.12}_{-0.08}(\phi_{f_1})
^{+0.01}_{-0.00}(a_t)^{+0.00}_{-0.01}(V)\;,
\eeq
\beq
R^{dd}_6[f_1(1285)/f_1(1420)] &\equiv& \frac{{\cal B}(B_d^0 \to f_1(1285) f_1(1420))_{\rm PQCD}}{{\cal B}(B_d^0 \to f_1(1420) f_1(1420))_{\rm PQCD}} \non
&\approx& 5.05^{+0.11}_{-0.19}(\omega_B)^{+0.54}_{-0.60}(f_{M})
^{+0.01}_{-0.38}(a_{f_1})^{+1.17}_{-1.08}(\phi_{f_1})
^{+0.04}_{-0.00}(a_t)^{+0.11}_{-0.09}(V)\;,
\eeq
\beq
R^{ss}_7[f_1(1420)/f_1(1285)] &\equiv& \frac{{\cal B}(B_s^0 \to f_1(1420) f_1(1420))_{\rm PQCD}}{{\cal B}(B_s^0 \to f_1(1285) f_1(1420))_{\rm PQCD}} \non
&\approx& 4.49^{+0.35}_{-0.45}(\omega_B)^{+0.02}_{-0.01}(f_{M})
^{+0.17}_{-0.39}(a_{f_1})^{+0.68}_{-0.42}(\phi_{f_1})
^{+0.14}_{-0.10}(a_t)^{+0.01}_{-0.00}(V)\;,
\eeq
\beq
R^{dd}_8[f_1(1285)/f_1(1420)] &\equiv& \frac{{\cal B}(B_d^0 \to f_1(1285) f_1(1285))_{\rm PQCD}}{{\cal B}(B_d^0 \to f_1(1420) f_1(1420))_{\rm PQCD}} \non
&\approx& 6.64^{+0.00}_{-0.10}(\omega_B)^{+1.79}_{-1.78}(f_{M})
^{+1.18}_{-0.78}(a_{f_1})^{+1.20}_{-1.26}(\phi_{f_1})
^{+0.72}_{-0.59}(a_t)^{+0.35}_{-0.31}(V)\;,
\eeq
\beq
R^{ss}_9[f_1(1420)/f_1(1285)] &\equiv& \frac{{\cal B}(B_s^0 \to f_1(1420) f_1(1420))_{\rm PQCD}}{{\cal B}(B_s^0 \to f_1(1285) f_1(1285))_{\rm PQCD}} \non
&\approx& 9.11^{+2.07}_{-1.79}(\omega_B)^{+3.05}_{-2.36}(f_{M})
^{+1.92}_{-2.74}(a_{f_1})^{+1.02}_{-1.18}(\phi_{f_1})
^{+0.31}_{-0.16}(a_t)^{+0.02}_{-0.02}(V)\;.
\eeq
Generally speaking, it should be noted that the errors arising from
the parameters, in particular, $f_{f_n}$ and $f_{f_s}$, and
$a_2^{||}$ and $a_1^\perp$, cannot be reduced
effectively in the ratios because of the significant interferences among the
decay amplitudes of $B_{d,s}^0 \to f_n f_n, f_n f_s,$ and $f_s f_s$.
Nevertheless,
we still expect that the LHCb and/or Belle-II experiments could perform
the measurements with enough precision on these ratios in the future,
in order to give some
essential constraints on the input parameters or the mixing angle $\phi_{f_1}$.

\begin{figure}[!!htb]
\begin{center}
\hspace{-1 cm}
\includegraphics[scale=0.55]{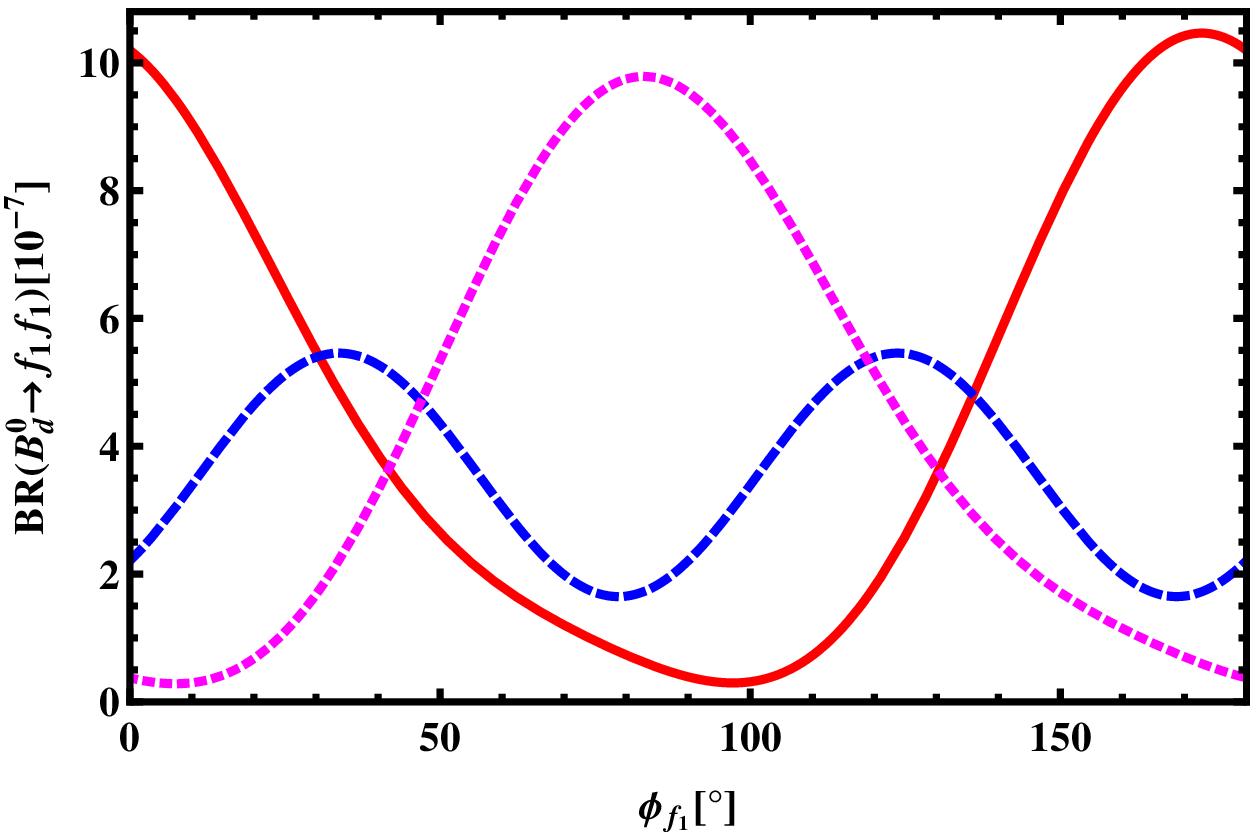}\hspace{1.2cm}
\includegraphics[scale=0.55]{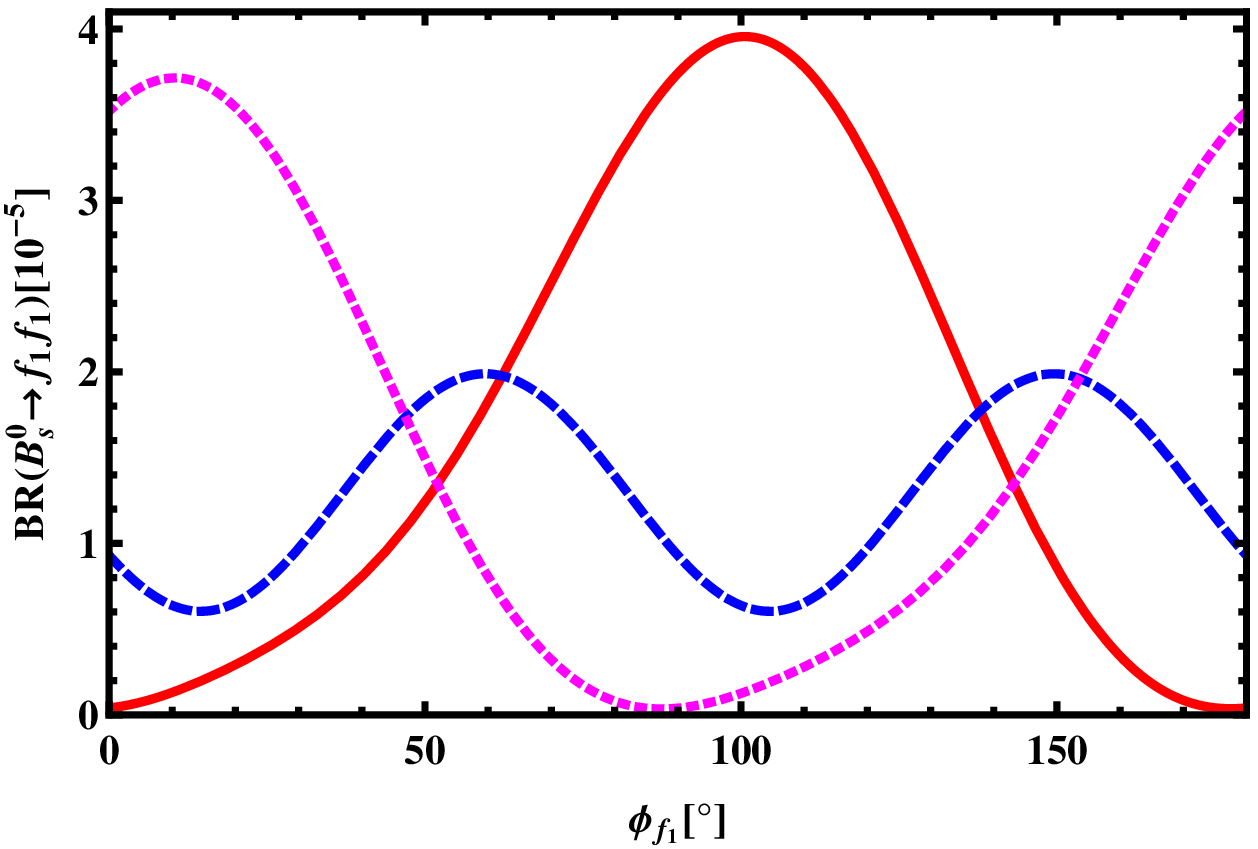}
\caption{(Color online) Dependence of the
{\it CP}-averaged $B_{d,s}^0 \to f_1 f_1$ branching ratios on
$\phi_{f_1}$ in the PQCD approach,
in which the red, solid line, the blue, dashed line, and the
magenta, dotted line correspond to the $B_{d,s}^0$ decays with final states
$f_1(1285) f_1(1285)$,
$f_1(1285) f_1(1420)$, and $f_1(1420) f_1(1420)$, respectively. }
\label{fig:fig2}
\end{center}
\end{figure}

From the numerical results for the decay amplitudes of the $B_{d,s}^0 \to f_1 f_1$
decays and the $B_{d,s}^0 \to f_n f_n, f_n f_s,$ and $f_s f_s$ flavor states
collected in the Tables~\ref{tab:DecAmp-d-f1f1}-\ref{tab:DecAmp-fnfs-ds},
one can find that the constructive or destructive interferences
with different extents in the
considered decays could vary with the mixing angle $\phi_{f_1}$
between $f_n$ and $f_s$ in the quark-flavor basis. To see the
variation clearly with the mixing angle, we show the {\it CP}-averaged
branching ratios ${\cal B}(B_{d,s}^0 \to f_1 f_1)$ varying with
$\phi_{f_1} \in [0, \pi]$ in Fig.~\ref{fig:fig2}.
It is of great interest to find that the line-shapes of ${\cal B}(B_d^0
\to f_1(1285) f_1(1285))$ and ${\cal B}(B_s^0 \to f_1(1420) f_1(1420))$
in the PQCD approach vary with $\phi_{f_1}$ similar to each other,
but with a quasi mirror symmetry, i.e., the
red, solid line in l.h.s and the magenta, dotted line in r.h.s, as shown in Fig.~\ref{fig:fig2}. The differences between these two lines are indeed induced by the dramatic interferences arising from the contributions from $B_{d}^0 \to f_n f_s$ to the former while that from $B_s^0 \to f_n f_s$ to the latter. And similar phenomena also
appear in the line shapes of ${\cal B}(B_d^0 \to f_1(1285) f_1(1420))$ and ${\cal B}(B_s^0 \to f_1(1285) f_1(1420))$, as well as of ${\cal B}(B_d^0 \to f_1(1420) f_1(1420))$ and ${\cal B}(B_s^0 \to f_1(1285) f_1(1285))$.
This picture really displays
the different interferences mainly from $B_d^0 \to f_n f_n$ and $B_d^0 \to f_n f_s$
to the $B_d^0 \to f_1 f_1$ decays while from $B_s^0 \to f_s f_s$ and $B_s^0 \to f_n f_s$ to the $B_s^0 \to f_1 f_1$ decays.
Of course, as given in the Table~\ref{tab:DecAmp-fnfs-ds}, the decay amplitudes from $B_d^0 \to f_s f_s$ and $B_s^0 \to f_n f_n$ in the longitudinal polarization also contribute to the $B_{d,s}^0 \to f_1 f_1$ decays. Therefore, it is expected that, if the precise $f_n$ and $f_s$ distribution amplitudes are available, then the mixing angle $\phi_{f_1}$ could be constrained by the near future measurements on the large $B_{s}^0 \to f_1 f_1$ decay rates associated with the interferences as exhibited in Fig.~\ref{fig:fig2}, and vice versa.

\subsection{{\it CP}-averaged polarization fractions}

Now, we turn to the calculations for the polarization fractions of the
$B_{d,s}^0 \to f_1 f_1$ decays.
Based on the helicity amplitudes $A_i(i=L, N, T)$, we can equivalently
define the amplitudes
in the transversity basis as follows:
\beq
{\cal A}_{L}&=& \xi m^{2}_{B_{d,s}^0} A_{L}, \qquad
{\cal A}_{\parallel}=\xi \sqrt{2}m^{2}_{B_{d,s}^0} A_{N}, \qquad
{\cal A}_{\perp}=\xi m_{B_{d,s}^0}^2 r^2_{f_1}
\sqrt{2(r^{2}-1)} A_{T}\;,
\label{eq:ase}
\eeq
for the longitudinal, parallel, and perpendicular polarizations,
respectively, with the normalization factor
$\xi=\sqrt{G^2_{F}P_c/(16\pi m^2_{B}\Gamma)}$ and the ratio
$r=P_{2}\cdot P_{3}/(m_{f_1}\cdot
m_{f_1})$. These amplitudes satisfy the relation,
\begin{eqnarray}
|{\cal A}_{L}|^2+|{\cal A}_{\parallel}|^2+|{\cal A}_{\perp}|^2=1 \;,
\end{eqnarray}
following the summation in Eq.~(\ref{dr1}).
Since the transverse-helicity contributions can manifest themselves through polarization observables, we therefore define {\it CP}-averaged fractions in three polarizations $f_{L}$, $f_\parallel$, and $f_\perp$ as the following,
\beq
f_{L,||,\perp}&\equiv& \frac{|{\cal
A}_{L,||,\perp}|^2}{|{\cal A}_L|^2+|{\cal A}_{||}|^2+|{\cal
A}_{\perp}|^2} = |{\cal
A}_{L,||,\perp}|^2.\label{eq:pf}
\eeq
With the above transversity amplitudes shown in Eq.~(\ref{eq:ase}), the relative phases
$\phi_{\parallel}$ and $\phi_{\perp}$ can be defined as
 \beq
 \phi_{\parallel} &=& \arg\frac{{\cal A}_{\parallel}}{{\cal A}_L} \;,
   \qquad
 \phi_{\perp} = \arg\frac{{\cal A}_{\perp}}{{\cal A}_L} \;.
 \eeq

From the Tables~\ref{tab:f12f14-ds}-\ref{tab:f14f14-ds}, one can clearly
find that four of the considered $B_{d,s}^0 \to f_1 f_1$ decays are
dominated by the transverse contributions, while the other two are governed
by the longitudinal ones, whose values for the polarization fractions $f_L$
and $f_T(=1-f_L)$ can be explicitly read as follows,
\beq
f_L(B_d^0 \to f_1(1285) f_1(1285)) &=& 27.7^{+19.7}_{-11.4}\%\;, \qquad
f_T(B_d^0 \to f_1(1285) f_1(1285)) = 72.3^{+8.4}_{-14.1}\% \;,
\label{eq:pf-22-d}\\
f_L(B_d^0 \to f_1(1285) f_1(1420)) &=& 18.0^{+14.1}_{-6.0}\%\;, \qquad
f_T(B_d^0 \to f_1(1285) f_1(1420)) = 81.9^{+4.6}_{-10.0}\%\;,
\label{eq:pf-24-d}\\
f_L(B_d^0 \to f_1(1420) f_1(1420)) &=& 12.4^{+21.3}_{-8.1}\% \;, \qquad
f_T(B_d^0 \to f_1(1420) f_1(1420)) = 87.6^{+6.0}_{-15.3}\%\;,
\label{eq:pf-44-d}\\
f_L(B_s^0 \to f_1(1420) f_1(1420)) &=& 10.6^{+10.3}_{-4.5}\% \;, \qquad
f_T(B_s^0 \to f_1(1420) f_1(1420)) = 89.4^{+3.3}_{-7.6}\%\;,
\label{eq:pf-44-s}
\eeq
and
\beq
f_L(B_s^0 \to f_1(1285) f_1(1285)) &=& 78.9^{+5.2}_{-7.5}\%\;, \qquad
f_T(B_s^0 \to f_1(1285) f_1(1285)) = 21.1^{+5.4}_{-3.8}\%\;,
\label{eq:pf-22-s}\\
f_L(B_s^0 \to f_1(1285) f_1(1420)) &=& 59.0^{+10.0}_{-13.3}\%\;, \qquad
f_T(B_s^0 \to f_1(1285) f_1(1420)) = 40.9^{+9.6}_{-7.0}\%\;,
\label{eq:pf-24-s}
\eeq
in which, all the errors from various parameters have been added in quadrature.
These predicted {\it CP}-averaged polarization fractions will be tested
at LHCb and/or Belle-II to further explore the decay mechanism with
helicities associated with experimental confirmations on the decay rates.

In Ref.~\cite{Cheng:2008gxa}, the longitudinal polarization fractions of
the $B_d^0 \to f_1 f_1$ decays
have been calculated in the QCDF approach,
\beq
f_L(B_d^0 \to f_1(1285) f_1(1285))_{\rm QCDF} &=& 0.66^{+0.07}_{-0.84}\;,\\
f_L(B_d^0 \to f_1(1285) f_1(1420))_{\rm QCDF} &=& 0.57^{+0.10}_{-0.66}\;,\\
f_L(B_d^0 \to f_1(1420) f_1(1420))_{\rm QCDF} &=& 0.68^{+0.23}_{-0.58}\;.
\eeq
As far as the central values are considered, the results in the QCDF approach
exhibit the dominance of the longitudinal decay amplitudes, which are very contrary
to those in the PQCD approach at leading order. However, when the
large uncertainties are taken into account, then one can find that the
transverse contributions can also govern these $B_d^0 \to f_1 f_1$ decays possibly.

In order to show explicitly the interferences among different flavor
states contributing to the $B_{d,s}^0 \to f_1 f_1$ decays in three polarizations, we collect their corresponding decay amplitudes, namely, $A_L(B_{d,s}^0 \to f_n f_n)$, $A_N(B_{d,s}^0 \to f_n f_s)$, and
$A_T(B_{d,s}^0 \to f_s f_s)$ in the Table~\ref{tab:DecAmp-fnfs-ds},
and the resultant amplitudes for the physical states,
i.e., $A_L(B_{d,s}^0 \to f_1 f_1)$, $A_N(B_{d,s}^0 \to f_1 f_1)$, and $A_T(B_{d,s}^0 \to
f_1 f_1)$, with differentiating them from tree diagrams and from penguin diagrams in the Tables~\ref{tab:DecAmp-d-f1f1} and \ref{tab:DecAmp-s-f1f1}. From these results quoted with only central values, one can easily observe that, generally speaking, the significantly
constructive [destructive] interferences govern the $B_{d}^0 \to f_1 f_1$ and $B_s^0
\to f_1(1420) f_1(1420)$ [$B_s^0 \to f_1(1285) f_1(1285)$] decays transversely [longitudinally]. While for the $B_s^0 \to f_1(1285) f_1(1420)$ mode, the interferences are slightly moderate within errors on both longitudinal and transverse
polarizations.

Unfortunately, no data or theoretical predictions for the considered $B_{s}^0 \to f_1 f_1$ decays are available nowadays. It is therefore expected that our predictions in the PQCD approach would be confronted with future LHCb and/or Belle-II experiments, as well as the theoretical comparisons within the framework of QCDF, soft-collinear effective theory~\cite{Bauer:2000yr}, and so forth.

Although, as aforementioned, the global agreement with
data for $B \to VV$ decays has been greatly improved
in the PQCD approach theoretically~\cite{Zou:2015iwa}
by picking up higher power $r_i^2$ terms
that were previously neglected, it seems that the predictions
about the polarization fractions for the $B_{d,s}^0
\to f_1 f_1$ decays cannot be understood similarly as the
$B_{d,s}^0 \to \omega \omega, \phi \phi$ decays~\cite{Zou:2015iwa}
due to the constructive and/or destructive
interferences with different extents.

Let us take the $B_{d,s}^0 \to f_1(1285) f_1(1285)$ and $B_{d,s}^0 \to f_1(1420) f_1(1420)$
decays as examples to clarify the differences from the vector decays of
$B_{d,s}^0 \to \omega \omega$ and $B_{d,s}^0 \to \phi \phi$ with ideal mixing
in the PQCD approach at leading order~\cite{Zou:2015iwa}. As we know,
with the referenced value $\phi_{f_1}=24^\circ$,
the physical states $f_1(1285)$ and $f_1(1420)$ are predominated by the
component of
$f_n$ and $f_s$ with a factor about $\cos\phi_{f_1} = 0.914$.
Thus, due to the similar behavior of the vector and
$^3\!P_1$-axial-vector mesons, the $B_{d,s}^0 \to
f_1(1285) f_1(1285)$ and $B_{d,s}^0 \to f_1(1420) f_1(1420)$ decays, in principle,
could provide
similar phenomena when the interferences from the $f_s$ and $f_n$ component are turned off correspondingly. Numerically, the $B_{d,s}^0 \to f_1(1285) f_1(1285)$
and $B_{d,s}^0 \to f_1(1420) f_1(1420)$ decay rates and the longitudinal polarization fractions without the relevant interferences
mentioned above could be read as,
\begin{itemize}
\item[]{(a) Without the interferences from the $f_s$ component, }
\beq
{\cal B}(B_d^0 \to f_1(1285) f_1(1285)) &= & 1.03 \times 10^{-6} \cdot \cos^2\phi_{f_1} \;, \qquad
f_L(B_d^0 \to f_1(1285) f_1(1285))= 17.5\%\;,\\
{\cal B}(B_s^0 \to f_1(1285) f_1(1285)) &=& 4.04 \times 10^{-7} \cdot \cos^2\phi_{f_1}\;, \qquad
f_L(B_s^0 \to f_1(1285) f_1(1285)) =100\%\;.
\eeq
\item[]{(b) Without the interferences from the $f_n$ component, }
\beq
{\cal B}(B_d^0 \to f_1(1420) f_1(1420)) &= & 0.37 \times 10^{-7} \cdot \cos^2\phi_{f_1} \;, \qquad
f_L(B_d^0 \to f_1(1420) f_1(1420))= 99.9\%\;,\\
{\cal B}(B_s^0 \to f_1(1420) f_1(1420)) &=& 3.52 \times 10^{-5} \cdot \cos^2\phi_{f_1}\;, \qquad
f_L(B_s^0 \to f_1(1420) f_1(1420)) =22.8\%\;.
\eeq
\end{itemize}
The ideal mixing, i.e., $\phi_{f_1}=0^\circ$, in the above equations would give the cases like $B_{d,s}^0 \to \omega \omega$ and $B_{d,s}^0 \to \phi \phi$ decays, because the $B_{d,s}^0 \to f_1(1285) f_1(1285)$
decays just receive the contributions from the flavor amplitude
$B_{d,s}^0 \to f_n f_n$ while the $B_{d,s}^0 \to f_1(1420) f_1(1420)$ ones just from the
$B_{d,s}^0 \to f_s f_s$ correspondingly.
Here, it is noted that,
in comparison to the decay constants $f_\omega = 0.187$~GeV, $f_\omega^T = 0.151$~GeV
and $f_\phi =0.215$~GeV, $f_\phi^T =0.186$~GeV for the vector $\omega$ and
$\phi$~\cite{Ball:2004rg},  the decay constants $f_{f_n} =0.193$~GeV
and $f_{f_s}=0.230$~GeV~\cite{Verma:2011yw} can
remarkably enhance the contributions on the transverse polarization
in the $B_{d}^0 \to f_n f_n$
and $B_{s}^0 \to f_s f_s$ decays, respectively, which finally result in the small longitudinal polarization fractions already presented in the above equations.
Meanwhile, one can easily observe that the interferences from the
$B_d^0 \to f_n f_s$ and $B_d^0 \to f_s f_s$ amplitudes contribute destructively to
the $B_d^0 \to f_1(1285) f_1(1285)$ decay rate with 23\%,
while those from the $B_s^0 \to f_n f_s$ and $B_s^0 \to f_n f_n$
amplitudes contribute constructively to the $B_s^0 \to f_1(1420) f_1(1420)$
branching ratio with 15\%. For the pure annihilation decays $B_d^0 \to f_s f_s$
and $B_s^0 \to f_n f_n$, as shown in the Table~\ref{tab:DecAmp-fnfs-ds},
both of them are absolutely governed by the longitudinal
contributions and the polarization fractions are about 100\%, which are almost the same as
those in the $B_d^0 \to \phi \phi$ and $B_s^0 \to \omega \omega$ decays correspondingly.

As presented in the Eqs.~(\ref{eq:fnfn-d})-(\ref{eq:amp-44-s}) and in the Tables~\ref{tab:DecAmp-d-f1f1}-\ref{tab:DecAmp-fnfs-ds}, one can find that,
except for the other four penguin-dominated modes,
the $B_d^0 \to f_1(1285) f_1(1285)$ and $B_d^0 \to f_1(1285) f_1(1420)$ decays
received the contributions arising from the color-suppressed tree amplitudes
with different extents.
To our knowledge, the $B_d^0 \to \rho^0 \rho^0$ decay dominated by the color-suppressed
tree amplitude has a small decay rate and a similarly small longitudinal polarization fraction in the PQCD approach at leading order that cannot be comparable to the
measurements. Nevertheless, the partial next-to-leading order
contributions from vertex corrections, quark loop, and
chromomagnetic penguin diagrams~\cite{Li:2006cva}, and the evolution from the
Glauber-gluon associated with the transverse-momentum-dependent wave functions~\cite{Liu:2015sra} could remarkably enhance the
branching ratio and the longitudinal polarization fractions simultaneously. Then the
theoretical predictions could be consistent well with the measurements given by
BABAR~\cite{Aubert:2008au} and LHCb~\cite{Aaij:2015ria} experiments within errors~\footnote{It should be mentioned that
previously the Belle collaboration reported a small longitudinal polarization
fraction of the $B_d^0 \to \rho^0 \rho^0$ decay~\cite{Adachi:2012cz} , which
is in good agreement with the values in the PQCD approach at leading order~\cite{Zou:2015iwa}, but is different dramatically to that
in the QCDF approach~\cite{Beneke:2006hg,Cheng:2008gxa,Cheng:2009cn} and the
soft-collinear effective theory~\cite{Wang:2017rmh}. It means that a refined
measurement on this small longitudinal polarization fraction is very important
at the Belle-II experiment. }. Therefore, the future examinations at LHCb
and/or Belle-II experiments on these two mentioned channels sensitive to the above-mentioned color-suppressed tree amplitudes could help to identify
the needs of the possible next-to-leading order corrections. Of course, this issue
has to be left for future study elsewhere.

In addition, we present the relative phases(in units of rad) $\phi_{\parallel}$
and $\phi_{\perp}$ of the $B_{d,s}^0 \to f_1 f_1$ decays for the first time.
The numerical results can be seen explicitly in the Tables~\ref{tab:f12f14-ds}-\ref{tab:f14f14-ds}.
Of course, these predictions have to await for the future tests because
no measurements have been available until now.

\subsection{Direct {\it CP}-violating asymmetries}

Now we come to the evaluations of direct {\it CP}-violating asymmetries
of the $B_{d,s}^0 \to f_1 f_1$ decays in the PQCD approach.
As for the direct {\it CP} violation
$\acp^{\rm dir}$, it is defined as
 \beq
\acp^{\rm dir} &\equiv& \frac{\overline{\Gamma} - \Gamma}{\overline{\Gamma} + \Gamma}
=  \frac{|\overline{ A}_{\rm final}|^2 - |{ A}_{\rm final}|^2}{
 |\overline{ A}_{\rm final}|^2+|{ A}_{\rm final}|^2},
\label{eq:acp1}
\eeq
where $\Gamma$ and ${ A}_{\rm final}$ stand for the decay rate and
the decay amplitude of $B_{d,s}^0 \to f_1 f_1$,
while $\overline{\Gamma}$ and $\ov{ A}_{\rm final}$ denote the
charge conjugation ones, correspondingly.
Meanwhile, according to Ref.~\cite{Beneke:2006hg}, the
direct-induced {\it CP} asymmetries
can also be studied with the help of helicity amplitudes.
Usually, we need to combine three polarization fractions, as
shown in Eq.~(\ref{eq:pf}), with those corresponding conjugation
ones of $B$ decays and then to quote the resultant six observables
to define direct {\it CP} violations of $B_{d,s}^0 \to f_1 f_1$ decays
in the transversity basis as follows:
\beq
\acp^{\rm dir,\ell}&=&
\frac{\bar f_\ell- f_\ell}{\bar f_\ell+
f_\ell}\;,
\eeq
where $\ell=L,\parallel,\perp$ and the definition of
$\bar f$ is the same as that in~Eq.(\ref{eq:pf}) but for the
corresponding $\bar B_{d,s}^0$ decays.

Using Eq.~(\ref{eq:acp1}), we calculate the
direct {\it CP}-violating asymmetries
in the $B_{d,s}^0 \to f_1 f_1$ decays and present the results as shown in
Tables~\ref{tab:f12f14-ds}-\ref{tab:f14f14-ds}. Based on these numerical
values, some comments are in order:
\begin{enumerate}
\item[(1)]
The direct {\it CP}-violating asymmetries within still large theoretical errors
for the $B_{d,s}^0 \to f_1 f_1$ decays could be read straightforwardly
from the Tables~\ref{tab:f12f14-ds}-\ref{tab:f14f14-ds} as follows,
\beq
\acp^{\rm dir}(B_d^0 \to f_1(1285) f_1(1285)) &=& 26.5^{+14.0}_{-19.9}\%\;, \qquad
\acp^{\rm dir}(B_s^0 \to f_1(1285) f_1(1285)) = -0.3^{+1.3}_{-1.3}\% \;,
\label{eq:dcp-22-ds} \\
\acp^{\rm dir}(B_d^0 \to f_1(1285) f_1(1420)) &=& 54.7^{+4.4}_{-10.1}\%\;, \qquad
\acp^{\rm dir}(B_s^0 \to f_1(1285) f_1(1420)) = 2.8^{+0.8}_{-0.9}\%\;,
\label{eq:dcp-24-ds} \\
\acp^{\rm dir}(B_d^0 \to f_1(1420) f_1(1420)) &=& 25.4^{+10.6}_{-15.8}\%\;, \qquad
\acp^{\rm dir}(B_s^0 \to f_1(1420) f_1(1420)) = -1.9^{+0.6}_{-0.8}\%\;.
\label{eq:dcp-44-ds}
\eeq
where all the errors from various parameters as specified previously have been
added in quadrature. For the former decays with $\Delta S =0$, as exhibited in
the Table~\ref{tab:DecAmp-d-f1f1}, the considerable
tree or penguin contaminations lead to the large direct {\it CP} asymmetries.
While for the latter decays with $\Delta S =1$, as displayed in the Table~\ref{tab:DecAmp-s-f1f1}, the negligible tree pollution result in the
very small direct {\it CP} violations.
Currently, all these direct
{\it CP} violations seem hard to be detected in the near future
experimentally due to the small
decay rates for the former decays and due to the tiny {\it CP} asymmetries for the latter ones.

\item[(2)]
For the $B_s^0 \to f_1(1285) f_1(1285)$ and
$B_s^0 \to f_1(1285) f_1(1420)$ decays,
there exist the large direct {\it CP}-violating asymmetries in both
transverse polarizations, namely, parallel and perpendicular, with still
large theoretical errors as follows,
\beq
\acp^{\rm dir, ||}(B_s^0 \to f_1(1285) f_1(1285)) &=& 9.1^{+7.5}_{-6.6}\% \;, \qquad
\acp^{\rm dir, \perp}(B_s^0 \to f_1(1285) f_1(1285)) = 8.6^{+7.9}_{-6.7}\% \;,
\eeq
and
\beq
\acp^{\rm dir, ||}(B_s^0 \to f_1(1285) f_1(1420)) &=& 13.6^{+5.0}_{-5.8}\% \;, \qquad
\acp^{\rm dir, \perp}(B_s^0 \to f_1(1285) f_1(1420)) = 14.8^{+5.5}_{-6.4}\% \;,
\eeq
which may be easily accessible associated with the large decay rates in the order of $10^{-6} \sim
10^{-5}$ within theoretical errors. Other predictions about the direct {\it CP}
violations in every polarization for the considered $B_{d,s}^0 \to f_1 f_1$ decays
could be found out in the Tables~\ref{tab:f12f14-ds}-\ref{tab:f14f14-ds} explicitly,
we here will not list them individually. These results
could be tested in the (near) future at LHCb, Belle-II, and
other facilities such as the Circular Electron-Positron Collider.
\end{enumerate}

\bigskip

At last, we shall give some remarks on the important annihilation contributions~\footnote{
In principle, as part of the power corrections,
the nonfactorizable emission diagrams could also contribute to the observables.
However, associated with the symmetric behavior from the leading-twist distribution
amplitudes of the $f_n$ and $f_s$ states, the mentioned contributions are
much smaller than those from the annihilation diagrams due to the cancelation between
the two nonfactorizable emission diagrams, e.g., see Fig.~\ref{fig:fig1}(c) with hard gluon from valence antiquark and
\ref{fig:fig1}(d) with hard gluon from valence quark, respectively. Numerically,
by taking the $B_s^0 \to f_1(1420) f_1(1420)$ decay rate
${\cal B}(B_s^0 \to f_1(1420) f_1(1420)) \approx 3.37^{+3.27}_{-2.18}\times 10^{-5}$ as an example,
it is found that ${\cal B}(B_s^0 \to f_1(1420) f_1(1420)) \approx 0.65^{+0.71}_{-0.43}\times 10^{-5}$ without the contributions arising from the annihilation diagrams and ${\cal B}(B_s^0 \to f_1(1420) f_1(1420)) \approx 0.01^{+0.04}_{-0.01} \times 10^{-5}$ without the nonfactorizable emission and annihilation contributions. Therefore,
we here emphasize the more important power
corrections, i.e., annihilation contributions, in this paper.}.
In particular, the penguin annihilation contributions was proposed to explain
the polarization anomaly
in the $B \to \phi K^*$ decays in standard model~\cite{Kagan:2004uw}.
Subsequently, more studies about the $B \to VV$ decays based on the rich data
were made in a systematic manner, and the penguin-dominated channels were further
found to need the annihilation contributions to a great extent~\cite{Beneke:2006hg,Wang:2017rmh,Cheng:2008gxa,Cheng:2009cn,Cheng:2009xz,Zou:2015iwa}.
It is worth pointing out that, because of similar behavior between the vector and the $^3\!P_1$-axial-vector mesons, the authors proposed the similar annihilation
contributions, as in the $B \to VV$ decays, to estimate the $B \to A(^3\!P_1)V$ and
$A(^3\!P_1)A(^3\!P_1)$ decay rates and polarization fractions~\cite{Cheng:2008gxa}.

Therefore, we shall explore the important contributions from weak annihilation diagrams to the $B_{d,s}^0 \to f_1 f_1$ decays considered in this work. In order to clearly examine the important annihilation contributions, we present the explicit
decay amplitudes in the Tables~\ref{tab:DecAmp-d-f1f1} and \ref{tab:DecAmp-s-f1f1}
decomposed as tree diagrams and penguin diagrams
with and without annihilation contributions in three polarizations. To
show the variations of the considered $B_{d,s}^0 \to f_1 f_1$ decays
with no inclusion of the contributions from annihilation diagrams, we
shall list the observables numerically such as the {\it CP}-averaged
branching ratios, the polarization fractions, and the direct {\it CP}-violating
asymmetries by taking only the factorizable emission plus the non-factorizable emission decay amplitudes into account in the PQCD approach.

\begin{itemize}
\item{Branching ratios}

Without the contributions from annihilation diagrams, then the branching ratios
will turn out to be,
\beq
{\cal B}(B_d^0 \to f_1(1285) f_1(1285)) &=& 5.00^{+6.72}_{-3.38} \times 10^{-7}\;, \qquad
{\cal B}(B_s^0 \to f_1(1285) f_1(1285)) = 2.60^{+2.63}_{-1.85} \times 10^{-6}\;,
\label{eq:BR-22-ds-na}\\
{\cal B}(B_d^0 \to f_1(1285) f_1(1420)) &=& 3.72^{+4.55}_{-2.61} \times 10^{-7}\;, \qquad
{\cal B}(B_s^0 \to f_1(1285) f_1(1420)) = 4.83^{+4.09}_{-3.05} \times 10^{-6}\;,
\label{eq:BR-24-ds-na}\\
{\cal B}(B_d^0 \to f_1(1420) f_1(1420)) &=& 0.95^{+1.07}_{-0.70} \times 10^{-7}\;, \qquad
{\cal B}(B_s^0 \to f_1(1420) f_1(1420)) = 0.65^{+0.71}_{-0.43} \times 10^{-5}\;.
\label{eq:BR-44-ds-na}
\eeq

\item{Longitudinal polarization fractions}

By neglecting the weak annihilation contributions, the {\it CP}-averaged longitudinal polarization fractions of the $B_{d,s}^0 \to f_1 f_1$ decays
are written as,
\beq
f_L(B_d^0 \to f_1(1285) f_1(1285)) &=& 16.7^{+11.6}_{-4.3} \%  \;, \qquad
f_L(B_s^0 \to f_1(1285) f_1(1285)) = 45.8^{+8.1}_{-11.3}\% \;,
\label{eq:pf-22-na-ds}\\
f_L(B_d^0 \to f_1(1285) f_1(1420)) &=& 13.3^{+7.3}_{-2.2}\% \;, \qquad
f_L(B_s^0 \to f_1(1285) f_1(1420)) =   74.6^{+12.8}_{-13.9}\% \;,
\label{eq:pf-24-na-ds}\\
f_L(B_d^0 \to f_1(1420) f_1(1420)) &=& 21.0^{+4.9}_{-4.8}\%\;, \qquad
f_L(B_s^0 \to f_1(1420) f_1(1420)) = 6.0^{+16.8}_{-6.3}\% \;.
\label{eq:pf-44-na-ds}
\eeq

\item{Direct {\it CP}-violating asymmetries}

Without the contributions arising from annihilation type diagrams, the direct {\it CP}-violating asymmetries are then given as,
\beq
\acp^{\rm dir}(B_d^0 \to f_1(1285) f_1(1285)) &=&  -1.2^{{+12.3}}_{{-18.4}}\% \;, \qquad
\acp^{\rm dir}(B_s^0 \to f_1(1285) f_1(1285)) = -4.7^{+1.6}_{-2.6}\% \;,
\label{eq:dcp-22-na-ds}\\
\acp^{\rm dir}(B_d^0 \to f_1(1285) f_1(1420)) &=&  39.1^{+1.9}_{-5.3}\%\;, \qquad
\acp^{\rm dir}(B_s^0 \to f_1(1285) f_1(1420)) = -3.7^{+2.0}_{-3.2}\% \;,
\label{eq:dcp-24-na-ds}\\
\acp^{\rm dir}(B_d^0 \to f_1(1420) f_1(1420)) &=& 29.8^{+6.7}_{-8.1}\% \;, \qquad
\acp^{\rm dir}(B_s^0 \to f_1(1420) f_1(1420)) = -2.4^{+0.5}_{-1.6}\%\;.
\label{eq:dcp-44-na-ds}
\eeq
\end{itemize}
In the above equations, all the errors from various parameters have been
added in quadrature.
Generally speaking, within the still large theoretical errors, for the
branching ratios for example, by combining the results as presented in the
Eqs.~(\ref{eq:BR-22-ds})-(\ref{eq:BR-44-ds}), it seems that the numerical results
with and without the important annihilation contributions could be consistent with each
other in a $2\sigma$ standard deviation. However,
in light of the central values about the observables
for the $B_{d,s}^0 \to f_1 f_1$ decays, the results collected in the Eqs.~(\ref{eq:BR-22-ds})-(\ref{eq:BR-44-ds}),
~(\ref{eq:pf-22-d})-(\ref{eq:pf-24-s}),~(\ref{eq:dcp-22-ds})
-(\ref{eq:dcp-44-ds}), and~(\ref{eq:BR-22-ds-na})
-(\ref{eq:dcp-44-na-ds}) clearly show that
\begin{itemize}
\item[]{(a)}
the annihilation diagrams can contribute to the
$B_{d,s}^0 \to f_1 f_1$ decay rates with different ratios from the least 5\% to
the largest 80\%. Specifically, once the annihilation contributions are turned
off, then the {\it CP}-averaged branching ratios of the $B_{d,s}^0 \to f_1 f_1$
decays will decrease about 25\% for the $B_{d}^0 \to f_1(1285) f_1(1285)$ and $B_d^0 \to f_1(1285) f_1(1420)$ modes; and reduce around 30\% and 35\% for the $B_s^0 \to f_1(1285) f_1(1285)$ and $B_s^0 \to f_1(1285) f_1(1420)$ ones, respectively.
The annihilation diagrams can enhance ${\cal B}(B_s^0 \to f_1(1420) f_1(1420))$ from $0.65\times 10^{-5}$ to $3.37\times 10^{-5}$.

\item[]{(b)}
Indeed, the annihilation diagrams can modify the polarization fractions of the $B_{d,s}^0 \to f_1 f_1$ decays with different extents. Without the contributions from the annihilation diagrams, it is found that the $B_s^0 \to f_1(1285) f_1(1420)$ channel remains longitudinal polarization dominated but with a 26\% enhancement, the $B_d^0 \to f_1(1285) f_1(1285)$, $B_d^0 \to f_1(1285) f_1(1420)$, $B_s^0 \to f_1(1420) f_1(1420)$, and $B_d^0 \to f_1(1420) f_1(1420)$ decays remains transverse polarization dominated but with a 40\%, 26\%, 42\% reduction of $f_L$, and a near 70\% enhancement of $f_L$, respectively, and the $B_s^0 \to f_1(1285) f_1(1285)$ mode goes from a large longitudinal polarization fraction to a slightly larger transverse one than one half.

\item[]{(c)}
As claimed in~\cite{Chay:2007ep}, the annihilation diagrams in the heavy $B$ meson decays could contribute a large imaginary part,
as shown in the Tables~\ref{tab:DecAmp-d-f1f1} and \ref{tab:DecAmp-s-f1f1}, and act as the main source of large strong phase in the PQCD approach. Therefore,
the absence of the contributions from annihilation diagrams change the interferences highly between the weak and strong phases in the $B_{d,s}^0 \to f_1 f_1$ decays and finally results in the significant variations of the direct {\it CP}-violating asymmetries, even
the positive or negative signs.
\end{itemize}

\bigskip
\section{Conclusions and Summary} \label{sec:summary}

In short, we have analyzed the $B_{s}^0 \to f_1 f_1$ decays for the first time in the quark-flavor basis with the PQCD approach.
We obtained the small decay rates that are hard to be measured in the CKM suppressed
$B_d^0 \to f_1 f_1$ decays while the large branching ratios that are easy to be accessible in the CKM favored $B_s^0 \to f_1 f_1$ ones due to the interferences with
different extents among the flavor decay amplitudes $B_{d,s}^0 \to f_n f_n$,
$f_n f_s$, and $f_s f_s$. Particularly, the $B_s^0 \to f_1(1420) f_1(1420)$ decay
with a large branching ratio in the order of $10^{-5}$
is expected to be measured
through the $B_s^0 \to (K_S^0 K^\pm \pi^\mp)_{f_1(1420)}
(K_S^0 K^\pm \pi^\mp)_{f_1(1420)}$ channel. Our numerical results of
the observables such as the {\it CP}-averaged branching ratios, the polarization fractions, and the direct {\it CP}-violating asymmetries
indicate that the weak annihilation diagrams play
important roles in understanding the dynamics in these $B_{d,s}^0 \to f_1 f_1$ decays
in the PQCD approach.
Of course, these predictions in the PQCD approach await for the confirmations
from the future examinations, which could help us to
understand the annihilation decay mechanism in vector-vector
and vector-axial-vector $B$ decays in depth.
We explored the dependence of
${\cal B}(B_{d,s}^0 \to f_1 f_1)$ on the mixing angle $\phi_{f_1}$ in the quark-flavor basis and found the interesting line shapes to hint useful information.
In light of the large theoretical errors induced by the unconstrained inputs,
we also defined nine ratios of the $B_{d,s}^0
\to f_1 f_1$ decay rates to await for the (near) future measurements
at LHCb and/or Belle-II, even other facilities, e.g.,
Circular Electron-Positron Collider. Note that
the large uncertainties of the predicted branching ratios are canceled to a large extent in several ratios. Then the mixing angle $\phi_{f_1}$ between the flavor
states $f_n$ and $f_s$ could be further constrained, which would finally
help pin down the $\theta_{K_1}$ angle to understand the properties of
the light axial-vector mesons more precisely.


\begin{acknowledgments}
The authors thank Hai-Yang~Cheng and Ju-Jun~Xie for helpful discussions.
This work is supported in part by the National Natural Science
Foundation of China under Grants No.~11765012, No.~11775117, No.~11705159 and No.~11975195,
by the Qing Lan Project of Jiangsu Province under Grant No.~9212218405,
by the Natural Science Foundation of Shandong province under the Grants
No.~ZR2018JL001 and No.~ZR2019JQ04, by the Project of Shandong Province Higher Educational Science and Technology Program under Grant No.~2019KJJ007,
and by the Research Fund of Jiangsu Normal University under Grant No.~HB2016004.
Z.J. is supported by Postgraduate Research $\&$ Practice Innovation Program
of Jiangsu Province(Grant No.~KYCX20\_2225).
\end{acknowledgments}


\begin{appendix}

\section{ Wave functions and distribution amplitudes}
\label{sec:app1}

The heavy $B$ meson is usually treated as a heavy-light system
and its light-cone wave function
can generally be defined as~\cite{Keum:2000ph,Lu:2002ny}
\beq
\Phi_{B} &=& \frac{i }{\sqrt{2N_c}} \left\{(\psl +m_{B})\gamma_5
 \phi_{B}(x, k_T) \right\}_{\alpha\beta}\;,
\label{eq:def-bq}
\eeq
where $\alpha,\beta$ are the color indices;
$P$ is the momentum of $B$ meson; $N_c$ is the color factor; and
$k_T$ is the intrinsic transverse momentum of the light quark in $B$ meson.
Recent developments on the $B$ meson wave function and its distribution
amplitude can refer to, e.g., the
Refs.~\cite{Bwf}.

The $B$ meson distribution amplitude in the impact $b$ (not to be confused with the heavy quark $b$. Here, $b$ is the conjugate space coordinate of transverse momentum $k_T$)
space has been proposed as~\cite{Keum:2000ph}
\beq
\phi_{B}(x,b)&=& N_{B}x^2(1-x)^2
\exp\left[-\frac{1}{2}\left(\frac{xm_{B}}{\omega_{B}}\right)^2
-\frac{\omega_{B}^2 b^2}{2}\right] \;,
\eeq
and widely
adopted, for example, in~\cite{Keum:2000ph,Ali:2007ff,Zou:2015iwa,
Liu:2014doa,Liu:2014dxa,Liu:2014jsa,Liu:2015sra,Li:2010nn}.
This $B$ meson distribution amplitude obeys the following normalization condition,
\beq
\int_0^1 dx \phi_{B}(x, b=0) &=& \frac{f_{B}}{2 \sqrt{2N_c}}\;,\label{eq:norm}
\eeq
where $f_B$ is the decay constant of the $B$ meson related to the normalization
factor $N_B$. For the $B_d^0$ meson,
the shape parameter $\omega_B$ was fixed at
$0.40$~GeV with $f_{B}= 0.19$~GeV and $N_B = 91.745$
by combining the rich data and plenties of PQCD calculations
on the observables of $B^+$ and $B_d^0$ mesons' decays~\cite{Keum:2000ph,Lu:2002ny}.
Here, the assumption of isospin symmetry has been made.
For the $B_s^0$ meson, a somewhat larger momentum fraction
is adopted due to the heavier $s$ quark, relative to
the lightest $u$ or $d$ quark in the $B^+$ or $B_d^0$ mesons.
Therefore, by considering a small SU(3) symmetry-breaking effect,
we adopt the shape parameter $\omega_{B} = 0.50$~GeV with $f_{B} = 0.23$~GeV
for the $B_s^0$ meson~\cite{Ali:2007ff}, and the corresponding normalization
constant is $N_{B} = 63.67$.
In order to estimate the theoretical uncertainties induced
by the shape parameters, we also consider varying the shape parameter $\omega_{B}$
by 10\%, that is, $\omega_B = 0.40 \pm 0.04$~GeV for $B_d^0$ meson and $\omega_{B} = 0.50 \pm 0.05$~GeV for the $B_s^0$ meson, respectively.

The wave functions for the light flavor $f_{n(s)}$ state of the axial-vector
$f_1$ mesons can be written as~\cite{Yang:2007zt,Li:2009tx},
 \beq
\Phi^{L}_{f_{n(s)}} &=&  \frac{ 1}{\sqrt{2 N_c}}\gamma_5
  \biggl\{ m_{f_{n(s)}}\, {\epsl}_L \,\phi_{f_{n(s)}}(x)  +
 {\epsl}_L \, \psl\,\phi_{f_{n(s)}}^t(x)  + m_{f_{n(s)}}\, \phi_{f_{n(s)}}^s(x) \biggr\}_{\alpha\beta}\;,
 \eeq
 \beq
\Phi^{T}_{f_{n(s)}} &=&  \frac{ 1}{\sqrt{2 N_c}} \gamma_5
  \biggl\{ m_{f_{n(s)}}\, {\epsl}_T\, \phi_{f_{n(s)}}^v(x) +
{\epsl}_T\, \psl\, \phi_{f_{n(s)}}^T(x)+m_{f_{n(s)}}
i\epsilon_{\mu\nu\rho\sigma}\gamma_5\gamma^\mu {\epsl}_T^{\nu}
n^\rho v^\sigma \phi_{f_{n(s)}}^a(x) \biggr\}_{\alpha\beta}\;,
\eeq
for longitudinal and transverse polarizations, respectively,
with the polarization vectors $\epsilon_L$ and $\epsilon_T$ of
$f_{n(s)}$,
satisfying $P \cdot \epsilon=0$. $x$ denotes the momentum
fraction carried by quarks in $f_{n(s)}$,
 $n=(1,0,{\bf 0}_T)$
and $v=(0,1,{\bf 0}_T)$ are dimensionless lightlike unit vectors,
and $m_{f_{n(s)}}$ stands for the mass of light
axial-vector flavor state $f_{n(s)}$.
In addition, we adopt the convention $\epsilon^{0123}=1$ for the
Levi-Civita tensor $\epsilon^{\mu\nu\alpha\beta}$.

The twist-2 light cone distribution amplitudes
can generally be expanded as the Gegenbauer polynomials~\cite{Yang:2007zt}:
\beq
 \phi_{f_{n(s)}}(x)&=&\frac{f_{f_{n(s)}}}{2\sqrt{2N_c}} 6 x  (1-x) \left[ 1  + a_{2}^\parallel\, \frac{3}{2} ( 5(2x-1)^2  - 1 )
\right]\;,\\
\phi_{f_{n(s)}}^T(x)&=& \frac{f_{f_{n(s)}}}{2\sqrt{2N_c}}6 x (1-x)
\left[  3 a_{1}^\perp\, (2x-1) \right] \;,
\eeq

For the twist-3 ones, we use the following
form as in Ref.~\cite{Li:2009tx}:
\beq
\phi_{f_{n(s)}}^s(x)&=&\frac{f_{f_{n(s)}}}{4\sqrt{2N_c}} \frac{d}{dx}\Biggl[ 6x
(1-x) (  a_{1}^\perp (2x-1) )\Biggr]\;, \\
\phi_{f_{n(s)}}^t(x)&=&\frac{f_{f_{n(s)}}}{2\sqrt{2N_c}}\Biggl[
\frac{3}{2}\,a_{1}^\perp\,(2x-1) (3 (2x-1)^2-1)\Biggr],
\eeq
\beq
\phi_{f_{n(s)}}^v(x)&=&\frac{f_{f_{n(s)}}}{2\sqrt{2N_c}}
\Biggl[\frac{3}{4}(1+(2x-1)^2) \Biggr]\;,\\
\phi_{f_{n(s)}}^a(x)&=&\frac{f_{f_{n(s)}}}{8\sqrt{2N_c}}
\frac{d}{dx}\Biggl[6 x (1-x) \Biggr]\;.
\eeq
where $f_{f_{n(s)}}$ is the ``normalization" constant for the
flavor state $f_{n(s)}$ on both longitudinal and transverse
polarizations,
and
the Gegenbauer moments $a_{2}^{\parallel}$ and $a_{1}^{\perp}$
are as follows,
\beq
a_2^\parallel &=& \left\{ \begin{array}{ll}
-0.02^{+0.02}_{-0.02} \;\;\; (\;{\rm for}\; f_n\;),&  \\
-0.04^{+0.03}_{-0.03} \;\;\; (\;{\rm for}\; f_s\;),&   \\ \end{array} \right.
\qquad
a_1^\perp = \left\{ \begin{array}{ll}
-1.04^{+0.34}_{-0.34} \;\;\; (\;{\rm for}\; f_n\;),&  \\
-1.06^{+0.36}_{-0.36} \;\;\; (\;{\rm for}\; f_s\; ).&   \\ \end{array} \right.
\label{eq:Geng-fns}
\eeq

\end{appendix}


\end{document}